# Scaling methods for estimating macroscopic fundamental diagrams in urban networks with sparse stationary sensor coverage

Nandan Maiti *, Manon Seppecher, Ludovic Leclercq

*Univ. Eiffel, ENTPE, LICIT-ECO7, Lyon, 69500, France*

ARTICLE INFO

ABSTRACT

Accurately estimating traffic variables across unequipped portions of a network remains a significant challenge due to the limited amount of sensors-equipped links, such as loop detectors and probe vehicles. A common approach is to apply uniform scaling, treating unequipped links as equivalent to equipped ones, which leads to a strong bias in MFD estimation. Two alternative approaches are proposed: (1) Hierarchical network scaling and (2) Variogram-based data imputation. The hierarchical scaling method categorizes the network into several clusters according to spatial and functional characteristics, applying tailored scaling factors to each category. The variogram-based imputation leverages spatial correlations to estimate traffic variables for unequipped links, capturing spatial dependencies in urban road networks. Validation results show that the hierarchical scaling approach yields the most accurate estimates, demonstrating reliable performance with as little as 5% uniform detector coverage, while the variogram model fails to provide estimates. While the variogram-based method provides strong results with over 10% detector coverage, it is slightly less effective than the hierarchical scaling approach but performs better than the baseline non-hierarchical method.

## 1. Introduction

The Macroscopic Fundamental Diagrams (MFDs) have emerged as a critical tool in network traffic management, providing a relationship between aggregate traffic variables such as flow, density, and speed across an entire urban road network (Daganzo and Geroliminis, 2008). This concept offers valuable insights for network traffic control (Ampountolas et al., 2017; Sirmatel and Geroliminis, 2018; Sirmatel et al., 2021), particularly in perimeter control strategies (Aboudolas and Geroliminis, 2013; Mariotte and Leclercq, 2019; Sirmatel and Geroliminis, 2020, 2021; Jiang and Keyvan-Ekbatani, 2023; Kouvelas et al., 2023; Chen et al., 2024; Yu et al., 2025), where traffic inflow and outflow are managed to optimize network-wide performances. By monitoring traffic at a macroscopic level, the MFDs enable decision-makers to regulate traffic in real-time, improve congestion management, and enhance the overall efficiency of urban transportation systems.

Despite its potential, estimating reliable MFDs for a city network requires extensive empirical data. Loop detectors (LDs), the most commonly used data source for MFDs estimation, collect information from fixed sensors embedded in road infrastructure (Buisson and Ladier, 2009; Geroliminis and Sun, 2011a; Keyvan-Ekbatani et al., 2012; Saberi and Mahmassani, 2012; Aboudolas and Geroliminis, 2013; Keyvan-Ekbatani et al., 2013; Ambühl et al., 2021; Lee et al., 2023; Mousavizadeh and Keyvan-Ekbatani, 2024). However, LDs face limitations, such as positional biases in speed estimation (Leclercq et al., 2014; Maiti and Leclercq, 2025), and uneven distribution of sensors in a network (Lee et al., 2023), leading to incomplete coverage. To address these limitations, floating-car devices (FCD), or probe is often used alongside LDs to enhance spatial coverage, primarily for improving spatial speed estimation (Geroliminis and Daganzo, 2008; Gayah and Daganzo, 2011; Geroliminis and Sun, 2011b; Mahmassani et al., 2013; Tsubota et al., 2014; Ambühl and Menendez, 2016; Yang Beibei et al., 2016; Du et al., 2016; Saeedmanesh and Geroliminis, 2016; Ambühl et al., 2017; Dakic and Menendez, 2018; Mariotte et al., 2020a,b). However, FCD has its own challenges, including the unknown and variable penetration rate of vehicles equipped with GPS in time and space. Estimating network-wide traffic flows based on the assumption of homogeneous FCD penetration introduces uncertainty in the flow estimation, particularly when coverage is sparse (Leclercq et al., 2014; Shim et al., 2019; Fu et al., 2020; Jin et al., 2024).

* Corresponding author.
*E-mail address:* nandan.maiti@univ-eiffel.fr (N. Maiti).

The limitations of these data sources underscore a broader issue: the network-wide traffic data used for MFD estimation is typically incomplete. LDs and FCD are often available only for certain links in the network, resulting in what is known as an 'equipped network MFD', an MFD that only reflects the links where sensors or FCD are deployed (Mariotte et al., 2020a). Consequently, this equipped MFD may not be representative of the entire network. This is critical for simulation studies as the demand is usually given for the full region, and estimating capacity and other variables on the equipped network only creates a mismatch with trips that could totally or partially happen in the non-equipped network.

To scale the equipped network MFD to the full city network, previous studies have attempted to partition the network into homogeneous areas and apply a scaling factor that adjusts for the ratio between the total length of links in the network and the length of the equipped links, e.g., Mariotte et al. (2020a). While this approach can offer some level of approximation, it suffers from significant limitations. The partitioning of the network based on equipped network data can lead to erroneous estimations, as the unequipped links may exhibit different traffic characteristics (Jiang et al., 2023; Jiang and Keyvan-Ekbatani, 2023; Saeedmanesh and Geroliminis, 2016, 2017; Gu and Saberi, 2019; Johari et al., 2023). Furthermore, applying a uniform scaling factor across areas with varying types of links, such as arterials and local roads, may lead to inaccurate flow estimates.

There is a need for a more advanced approach to spatial scaling of the MFD. To overcome the challenges associated with uniform scaling, a new methodology that accounts for the link hierarchy within a network is proposed. This method involves developing separate scaling factors for different hierarchical levels of the network, such as major arterials, collectors, and local streets. By accounting for the differences in traffic behavior at various link levels, this approach offers a more accurate estimation of network flow variables, allowing for accurate scaling of MFD from sensor-equipped portions of the network to the entire city.

### 1.1. Literature review

#### 1.1.1. Equipped networks' MFDs

The literature about MFD estimation from 'equipped network' to 'full network' can be broadly classified into two groups based on the data used: one using empirical data sources from the equipped network, the other relying on large-scale simulation frameworks for a city. The main challenges associated with the empirical data are the limited coverage of LDs and FCD throughout the network, while the simulations fail to replicate the multimodal interactions and traffic dynamics of a complex network (Leclercq et al., 2014; Fu et al., 2020; Mariotte et al., 2020b; Paipuri et al., 2021). Thus, we mainly focused on the existing empirical study on MFD estimation and calibration.

MFD estimation predominantly relied on LDs data (Geroliminis and Daganzo, 2008; Buisson and Ladier, 2009) information on vehicle counts and occupancy rates on road segments, allowing the calculation of network-wide average flow and density. This approach was inspired by Edie (1963) definition of traffic flow metrics, where average flow and density are calculated based on the data collected from multiple equipped links within a network (Leclercq et al., 2014). However, the reliance on LD data introduces certain limitations inherent to the Eulerian observation framework, particularly the potential for bias in MFD estimation arising from the non-uniform spatial distribution of detectors (Leclercq et al., 2014). The placement of detectors, often in congested areas or near traffic signals, can result in an overestimation of network density (Lee et al., 2023; Maiti and Leclercq, 2025). Despite this limitation, studies have shown that MFDs derived from a small fraction of links can still approximate critical density ranges observed in the full network MFD (Keyvan-Ekbatani et al., 2013). Mariotte et al. (2020a) introduced the concept of an 'active network' to improve MFD estimation, focusing on major arterials and secondary roads. They highlighted the importance of scaling the production MFD based on the ratio of total network length to equipped network length. The estimated scaling factor for production is based on the assumption of homogeneous network partitioning (Saeedmanesh and Geroliminis, 2016, 2017; Jiang et al., 2023; Johari et al., 2023) using the equipped network information, leading to ambiguity in portioning for the full network (Johari et al., 2021). Saffari et al. (2020) expanded on this by investigating the selection of critical links for MFD estimation, though they noted that their findings were sensitive to the specific test bed used.

There are two main options for estimating the traffic states in non-equipped networks to derive complete MFDs. The first involves grouping non-equipped links with clusters of links exhibiting similar behavior and determining a scaling factor for each cluster. The second approach focuses on imputing missing link variables for non-equipped links. The following review outlines the existing studies for spatial scaling and imputation in network traffic state estimation.

#### 1.1.2. Spatial scaling

The main challenge with imputing traffic states at the unequipped links is the estimation errors due to a lack of available information from the surrounding links and heterogeneity of traffic states among links. Network traffic state estimation approaches can be broadly classified into two main categories: model-driven and data-driven (Seo et al., 2017; Takayasu et al., 2022; Saeedmanesh et al., 2021). High-accuracy model-driven methods often require additional information, such as detailed traffic signal settings or individual vehicle detector actuation data (Leclercq et al., 2014; Tilg et al., 2023). While these methods have proven efficient in estimating traffic states, particularly at the link level (Nantes et al., 2016; Maiti and Chilukuri, 2023b,a, 2024; Maiti et al., 2024), their accuracy is highly dependent on the quality of the embedded models. Data-driven approaches leverage large amounts of historical data to train models using machine learning or statistical techniques (Kong et al., 2009; Anuar et al., 2015; Lu et al., 2018; Takayasu et al., 2022; Maiti and Chilukuri, 2023b). Since data-driven methods do not rely on strict theoretical assumptions, the accuracy of the models is largely influenced by the quantity and quality of the data. However, missing data is a common issue in data-driven methods, and several studies have attempted to address this problem (Chen et al., 2003; Xu et al., 2015; Offor et al., 2019). Nowadays, graph convolution networks (GCNs), generative adversarial networks (GANs), and physics-informed neural networks (PINNs) are gaining significant attention in estimating the missing traffic flow variables in unequipped networks. GCNs struggle with capturing dynamic temporal changes in traffic networks, as they are often designed for static graphs. Additionally, the hierarchical and complex topology of traffic networks complicates model training (Rossi et al., 2021; Taguchi et al., 2021; Castro-Correa et al., 2023; Bao et al., 2023; Zhang et al., 2023). GANs face difficulties generating realistic traffic data due to the stochastic nature of traffic demand, making adversarial training unstable (Zhang et al., 2019; Zheng et al., 2022). PINNs, while incorporating physical laws, may struggle with noisy or incomplete data, as real-world traffic patterns can deviate significantly from theoretical models (Zhang et al., 2024; Usama et al., 2022). Balancing accuracy and computational efficiency remains a significant challenge across these approaches.

#### 1.1.3. *Geo-spatial imputation*

In transportation data, kriging has been used to impute missing values in Annual Average Daily Traffic (AADT) datasets, with studies showing its superiority over traditional regression models like ordinary least squares (OLS) (Eom et al., 2006; Wang and Kockelman, 2009). The kriging method outperformed OLS, particularly in regions with moderate-to-high traffic volumes. However, under low traffic flow conditions, it tends to overestimate the missing data (Wang and Kockelman, 2009). One of the limitations of the Kriging method is its reliance on Euclidean distance, which may not be ideal for transportation networks where distances are more appropriately measured along road networks. Zou et al. (2012) proposed the use of an approximated road network distance based on isometric embedding theory to address this issue, improving the interpolation accuracy, particularly in complex road networks. Nevertheless, studies such as Selby and Kockelman (2013) indicated that the use of road network distances does not always significantly improve prediction performance compared to traditional Euclidean distance-based methods.

In addition, while kriging has demonstrated success in spatial imputation, its application to spatio-temporal data remains relatively underexplored. Some studies, such as those by Yang et al. (2014), have extended kriging to account for both spatial and temporal dimensions. These efforts aim to better capture the spatio-temporal characteristics of traffic data. Moreover, Marcotte (1991) proposed cokriging, which uses secondary correlated variables alongside the primary variable, has shown promise in various fields. The application of cokriging for traffic data imputation is an area worth exploring, especially given the availability of multiple sources of traffic data (Shamo et al., 2015). A recent study by Bae et al. (2018) proposed the use of spatiotemporal cokriging to impute missing traffic flow speed data, integrating multiple data sources. Their results demonstrate that spatiotemporal cokriging improves the accuracy of imputation, particularly when missing data exhibit non-random patterns, such as blocks of missing data, which occur due to system malfunctions or maintenance issues. Later, Laval (2023) showed that the traffic flow variables such as travel times, relaxation times, and delay in urban networks near critical density follow the scaling with total network length.

Despite the advancements in using the MFD for traffic management, several gaps remain. Current methods rely on incomplete data from LDs and FCD, leading to inaccurate MFD estimates that do not fully represent the entire network. Uniform scaling approaches disregard the hierarchical nature of urban road networks, resulting in imprecise flow and density estimates across different road types. Additionally, advanced data-driven models like GCNs and PINNs struggle with missing or noisy data, limiting their effectiveness in estimating traffic states. Moreover, traditional spatial scaling techniques such as kriging are not well-explored for complex road networks. A more refined approach, incorporating link-hierarchical scaling and better data imputation, is needed to improve MFD accuracy for complete network estimation. This study develops a comprehensive framework for scaling network flow dynamics from equipped to full networks by incorporating road hierarchy and spatial imputation techniques. The main contributions of this study are listed below:

1. This work includes a novel hierarchical network scaling framework, which accounts for the distinct traffic characteristics of different road types, such as arterials and local streets, overcoming the inaccuracies of uniform scaling methods.
2. Utilizes geospatial imputation methods, such as variogram models, for network traffic data imputation. It leverages secondary data sources to fill gaps and inconsistencies in LDs coverage, ensuring a more accurate estimation of network traffic dynamics.
3. Provides validation of the proposed methodologies using real-world traffic data, demonstrating their effectiveness in capturing network-wide traffic dynamics.

## 2. Methodology

The proposed methodology estimates traffic variables for an entire network based on partially sensor-equipped data using hierarchical network scaling and variogram-based imputation approaches. Initially, the network is classified into equipped and unequipped links. In the hierarchical approach, links are categorized into several hierarchical levels based on spatial and functional characteristics, with tailored scaling factors applied to estimate traffic on unequipped links on each hierarchy. The spatial imputation approach employs variogram-based imputation to model spatial correlations between equipped and unequipped links, refining the traffic estimates. The outputs from these methods provide comprehensive estimates of traffic variables across the network, integrating observed data with imputed values for unequipped links. The results are benchmarked against a non-hierarchical uniform scaling method (Mariotte et al., 2020a), which applies consistent scaling factors across all links regardless of their spatial or functional characteristics. A summary of the methodological framework is provided in Fig. 1.

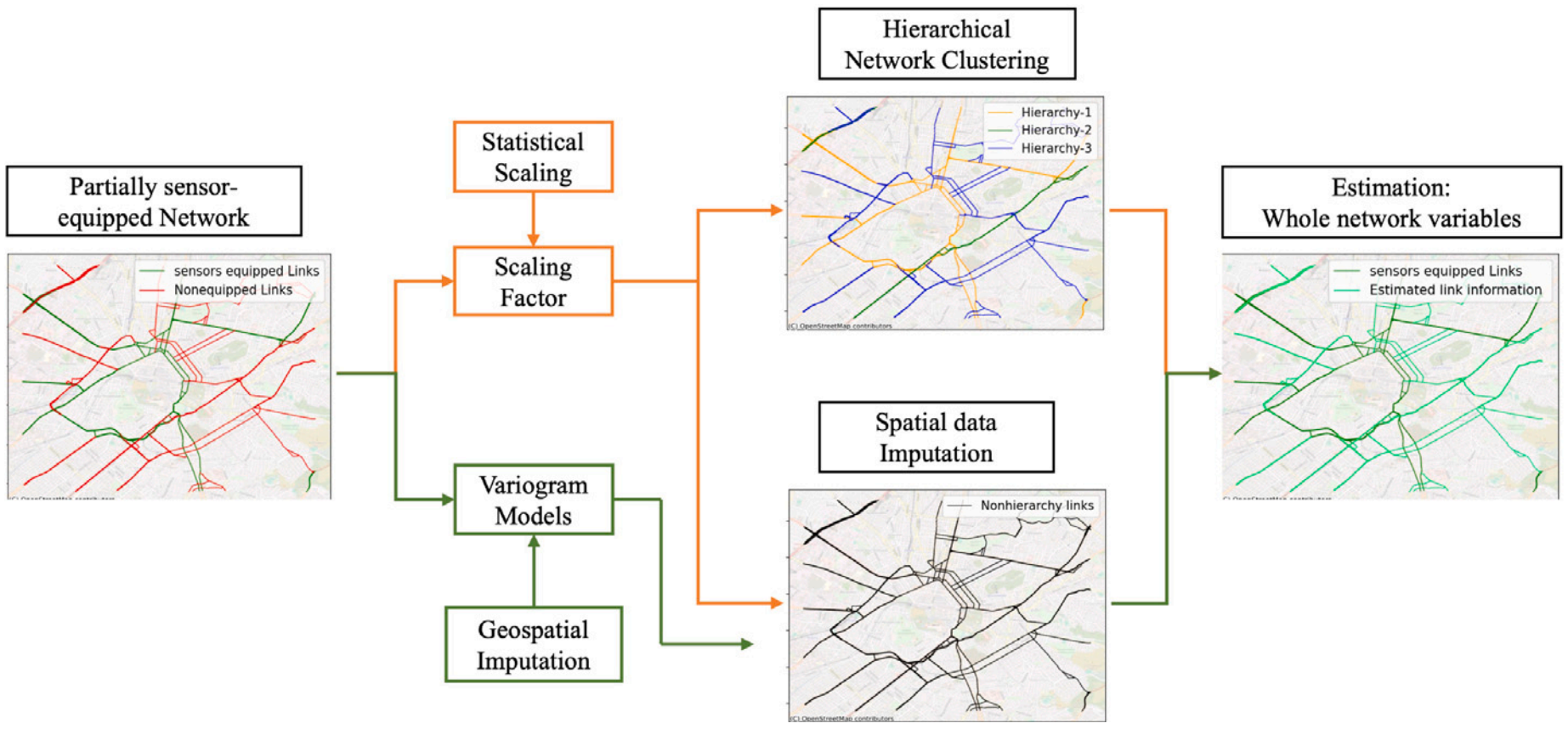

**Fig. 1.** An overview of the overall methodology for estimating whole network variables from the partially-equipped network.

### 2.1. Network variables estimation and scaling issues

#### 2.1.1. Uniform scaling or baseline

Let us assume a traffic network having links ($i$) of $n$ numbers ($|i| = n$), with a different hierarchy based on serviceability, defined as link class as $t$. Among all the links, some have LDs count as $j$. Therefore, the flow in the links of LDs can be presented as $\{q_j \mid j \in i\}$. The lengths of the links in the network are represented by $l_i$. The overall network flow ($\hat{q}_N$) can be expressed by total travel distance (TTD) as network 'production' by all vehicles over the time-space region, shown as follows in (1):

$$\hat{q}_N = \frac{\sum_{i=1}^{n} TTD_i}{\sum_{i=1}^{n} l_i \Delta t} \tag{1}$$

The TTD is the sum of equipped network TTD ($TTD_{j,eq}$) and non-equipped network TTD ($TTD_{i\notin i,neq}$)

$$TTD = \sum_{i=1}^{n} TTD_i = \sum_{\forall j\in\{eq\}} TTD_{j,eq} + \sum_{\forall i\in\{neq\}} TTD_{i,neq} \tag{2}$$

The first term of (2) can be calculated from the LDs.

$$\sum_{\forall j\in\{eq\}} TTD_{j,eq} = \sum_{\forall j\in\{eq\}} q_j l_j \tag{3}$$

The second term of (2) can be directly estimated after the imputation method. Otherwise, we know that the following expectation relation holds for all situations.

$$\mathbb{E}(q_i l_i) = \mathbb{E}(q_i)\mathbb{E}(l_i) + Cov(q_i, l_i) \tag{4a}$$

$$\frac{1}{p}\sum_{i\in\{neq\}} (q_i l_i) = \left(\frac{1}{p}\sum_{i\in\{neq\}} q_i\right)\frac{1}{p}\sum_{i\in\{neq\}} l_i + Cov_{neq}(q_i, l_i) \tag{4b}$$

$$\sum_{i\in\{neq\}} (q_i l_i) = \bar{q}_{neq}\sum_{i\in\{neq\}} l_i + p \times Cov_{neq}(q_i, l_i) \tag{4c}$$

Here, $\bar{q}_{neq}$ is the mean flow in the non-equipped network, and $p$ is the total number of non-equipped links ($|i \notin j| = p$). Also, assume $m$ is the total number of equipped links ($|j| = m$) in the network with mean flow estimated as $\bar{q}_{eq}$ from LDs. If we assume the non-equipped network shares the same average flow and covariance as the equipped, then we can calculate the second term of (2), as (5).

$$\sum_{\forall i\in\{neq\}} (q_i l_i) = \bar{q}_{neq}\sum_{\forall i\in\{neq\}} l_i + p \times Cov_{neq}(q_i, l_i) \tag{5a}$$

$$= \left(\frac{1}{m}\sum_{j\in\{eq\}} q_j\right)\sum_{\forall i\in\{neq\}} l_i + p \times Cov_{eq}(q_j, l_j) \tag{5b}$$

$$= \bar{q}_{eq}\sum_{\forall i\in\{neq\}} l_i + p \times Cov_{eq}(q_j, l_j) \tag{5c}$$

Note that the covariance term can be disregarded as it was proven negligible in all our data analyses. So, the TTD can be expressed as (6).

$$TTD = \bar{q}_{eq} \times \left( \sum_{\forall i \in \{neq\}} l_i + \sum_{\forall j \in \{eq\}} l_j \right) \tag{6a}$$

$$= \bar{q}_{eq} \times l_{net} \tag{6b}$$

Here $l_{net}$ is the total network length, including equipped and non-equipped links. The assumption, $\bar{q}_{neq} = \bar{q}_{eq}$ forms the foundation of the uniform scaling factor, where all variables are scaled by the factor of network length covered.

Unfortunately, this assumption is often invalid, and equipped links are not a randomly selected subset of the full network. We propose a hierarchical scaling approach that will cluster the equipped network into groups that more likely resemble parts of the non-equipped network.

### 2.1.2. Hierarchical network scaling

The general idea in the hierarchical network scaling method is that we want to approach $\bar{q}_{neq} = \bar{q}_{eq}$ at the cluster level. Similar to (5), we can use the $\bar{q}_{eq,t}$ to estimate the $\bar{q}_{neq,t}$, where $t$ represents link clusters. One option is to use the network hierarchy to define the clusters. Thus, we propose the following scaling method for the hierarchical network approach.

$$\sum_{\forall t} TTD_{eq,t} = \sum_{\forall j,t} q_{j,t} l_{j,t} \tag{7}$$

$$\mathbb{E}(q_{j,t} l_{j,t}) = \mathbb{E}(q_{j,t})\mathbb{E}(l_{j,t}) \tag{8}$$

$$\frac{1}{m} \sum_{\forall t} \sum_{j=1}^{m} q_{j,t} l_{j,t} = \sum_{\forall t} \left(\frac{1}{m} \sum_{j=1}^{m} q_{j,t}\right) \frac{1}{m} \sum_{\forall t} \sum_{j=1}^{m} l_{j,t} \tag{9}$$

$$\sum_{\forall j,t} q_{j,t} l_{j,t} = \sum_{\forall t} (\bar{q}_{eq,t}) \sum_{\forall j,t} l_{j,t} \tag{10}$$

Similarly, for non-equipped networks:

$$\sum_{\forall t} TTD_{neq,t} = \sum_{\forall i \in \{neq\}} q_{i,t} l_{i,t} \tag{11}$$

$$\mathbb{E}(q_{i,t} l_{i,t} | \forall i \notin j, t) = \mathbb{E}(q_{i,t} | \forall i \notin j, t)\mathbb{E}(l_{i,t} | \forall i \notin j, t) \tag{12}$$

$$\sum_{\forall t} \sum_{i=1}^{p} q_{i,t} l_{i,t} = \sum_{\forall t} (\bar{q}_{neq,t}) \sum_{\forall t} \sum_{i=1}^{p} l_{i,t} \tag{13}$$

$$\sum_{\forall i \notin j,t} q_{i,t} l_{i,t} = \sum_{\forall t} (\bar{q}_{neq,t}) \sum_{\forall i \notin j,t} l_{i,t} \tag{14}$$

Assuming the mean flow for a non-equipped network is the same as an equipped network for a unique hierarchy, i.e., ($\bar{q}_{neq,t} = \bar{q}_{eq,t}$). Therefore, by applying (10) to (14), we get the following estimation of TTD for a non-equipped network.

$$\sum_{\forall i \notin j,t} q_{i,t} l_{i,t} = \frac{\sum_{\forall j,t} q_{j,t} l_{j,t}}{\sum_{\forall j,t} l_{j,t}} \sum_{\forall i \notin j,t} l_{i,t} \tag{15}$$

$$= \sum_{\forall j,t} q_{j,t} l_{j,t} \frac{\sum_{\forall i \notin j,t} l_{i,t}}{\sum_{\forall j,t} l_{j,t}} \tag{16}$$

Thus, the total network flow can be estimated from the equipped network information as follows:

$$\hat{q} = \frac{\sum_{\forall t} TTD}{\sum_{\forall i,t} l_{i,t} \Delta t} \tag{17}$$

$$= \frac{\sum_{\forall t} TTD_{eq} + \sum_{\forall t} TTD_{neq}}{\sum_{\forall i,t} l_{i,t} \Delta t} \tag{18}$$

$$= \frac{\sum_{\forall j,t} q_{j,t} l_{j,t} + \left( \sum_{\forall j,t} q_{j,t} l_{j,t} \frac{\sum_{\forall i \notin j,t} l_{i,t}}{\sum_{\forall j,t} l_{j,t}} \right)}{\sum_{\forall i,t} l_{i,t} \Delta t} \tag{19}$$

Similarly, we can also derive the network average density from the equipped network. The network average density can be expressed by total travel time (TTT) as the 'accumulation' spent by all vehicles in the network over the time-space domain. Similar to the TTD in (2), we can estimate TTT in the equipped network as $TTT_{eq} = \sum_{\forall j} k_j l_j$. Local densities at LD-level ($k_j$) need to be corrected since the LDs suffer from location-biased and systematic errors. This study corrected LD-level density estimation as per the methods mentioned in Maiti and Leclercq (2025). Therefore, the total network density can be formulated from the corrected LDs in the equipped network as follows:

$$\hat{k} = \frac{\sum_{\forall t} TTT}{\sum_{\forall i,t} l_{i,t} \Delta t} \tag{20}$$

$$= \frac{\sum_{\forall t} TTT_{eq} + \sum_{\forall t} TTT_{neq}}{\sum_{\forall i,t} l_{i,t} \Delta t} \tag{21}$$

$$= \frac{\sum_{\forall j,t} k_{j,t} l_{j,t} + \left( \sum_{\forall j,t} k_{j,t} l_{j,t} \frac{\sum_{\forall i \notin j,t} l_{i,t}}{\sum_{\forall j,t} l_{j,t}} \right)}{\sum_{\forall i,t} l_{i,t} \Delta t} \tag{22}$$

The variables $q_{j,t}$ and $k_{j,t}$ represent flow and density in the equipped network of $t$ link hierarchy, estimated at the detector level on a link.

### 2.2. Spatial imputation: Variogram

Instead of scaling observations from the equipped network to represent the non-equipped network, an alternative approach is to impute flow and density values for all links and subsequently calculate the full MFDs. To achieve this, we propose a spatial variogram-based imputation method. Kriging is a well-established geostatistical interpolation method initially developed by Krige (1951) and later expanded by Matheron (1963). It has been applied across various disciplines for spatial data interpolation, including missing data imputation in traffic studies. The central component of kriging is the variogram, a function that models the spatial correlation between data points. The variogram provides an essential representation of how spatial dependence changes with distance, guiding Kriging's ability to estimate unknown values at unobserved locations based on nearby known observations (Cressie, 1993).

Variogram models, such as spherical, exponential, and Gaussian models, are typically fitted to the empirical variogram to quantify spatial continuity (Oliver, 2014). These models help define the structure of spatial relationships and enable kriging to make predictions.

#### 2.2.1. Spatial dependency and variogram

Let the traffic flow data for certain links in a network be given as $\{(s_i, q_i)\}$, where $s_i = (x_i, y_i)$ represents the geographical coordinates (latitude and longitude) of the $i$th known link, and $q_i$ is the observed traffic flow at that link. The task is to estimate traffic flow $q(s_0)$ at unknown locations $s_0$ by leveraging spatial interpolation techniques based on a variogram-based model.

The variogram measures the spatial dependence of a random variable, in this case, traffic flow, across a network. The variogram $\Gamma(h)$ represents how traffic flow differences are expected to change with increasing distance ($h$) between two locations. Unlike the traditional variogram model's spatial distance, in this study, the distance between two locations is measured by the shortest path distance along the road network. For any two points $s_i$ and $s_j$ separated by distance $h = \|s_i - s_j\|$, the variogram is defined as:

$$\Gamma(h) = \frac{1}{2} \mathbb{E} \left[ (q(s_i) - q(s_j))^2 \right] \tag{23}$$

Here, $\mathbb{E}$ denotes the expectation, and $h$ is the shortest path distance between $s_i$ and $s_j$.

The variogram is often approximated from data using the empirical variogram:

$$\Gamma(h) = \frac{1}{2N(h)} \sum_{s_i, s_j : \|s_i - s_j\| = h} (q_i - q_j)^2 \tag{24}$$

where $N(h)$ is the number of pairs of points separated by distance $h$. This empirical variogram helps identify the spatial structure in the traffic flow data. The empirical variogram, $\Gamma(h)$, is typically modeled using one of several functional forms (e.g., spherical, exponential, Gaussian) based on the empirical semivariances derived from the data. For instance, the spherical variogram model is expressed as:

$$\gamma(h) = \begin{cases} C_0 + C \left( \frac{3h}{2a} - \frac{h^3}{2a^3} \right), & \text{if } h \le a \\ C_0 + C, & \text{if } h > a \end{cases} \tag{25}$$

where $a$ is the range, $C_0$ is the nugget, and $C$ is the sill, serve as constants for the variogram model $\gamma(h)$.

#### 2.2.2. Spatial interpolation: Kriging system setup

Once the variogram model is fitted, the traffic flow at unmeasured locations can be estimated using *Ordinary Kriging*. In *Ordinary Kriging* for traffic flow estimation, we aim to predict the traffic flow $\hat{q}(s_0)$ at an unmeasured location $s_0$, using a weighted linear combination of the traffic flows $q(s_i)$ at known locations $s_i$, where $i = 1, 2, \dots, n$. The primary assumption in *Ordinary Kriging* is that the mean traffic flow $\bar{q}(s)$ is constant within a neighborhood of the target point $s_0$. The following assumptions are considered:

- The deterministic component $\bar{q}(s)$, representing the mean traffic flow, is approximately constant across the neighborhood of interest, so $\bar{q}(s_0) \approx \bar{q}(s_i) \equiv \bar{q}$.
- The stochastic component $\epsilon(s)$, representing random fluctuations in traffic flow, is Gaussian-distributed with zero mean, and the correlation between flows at different locations depends on their spatial separation.

The predicted traffic flow $\hat{q}(s_0)$ at location $s_0$ is given by:

$$\hat{q}(s_0) = \bar{q}(s_0) + \epsilon(s_0) \tag{26}$$

The random component can be defined as a weighted multiplication of the deviation of the observed traffic flow at $s_i$ from the mean $s_0$. The kriging weights $w_i$ are applied to these deviations, meaning that locations closer to $s_0$ (with higher spatial correlation) will have a greater influence on the prediction. This spatially weighted adjustment ensures that the predicted traffic flow at $s_0$ not only considers the global mean $\bar{q}$ but also incorporates the local variations (the deviations from the mean) based on the observed traffic flows at the nearby locations.

$$\hat{q}(s_0) = \bar{q} + \sum_{i=1}^{n} w_i \left( q(s_i) - \bar{q} \right) \tag{27}$$

Since $\bar{q}$ is assumed to be constant, the estimator simplifies to:

$$\hat{q}(s_0) = \sum_{i=1}^{n} w_i q(s_i) + \bar{q} \left( 1 - \sum_{i=1}^{n} w_i \right) \tag{28}$$

For an unbiased estimator, the weights $w_i$ must satisfy the constraint:

$$\sum_{i=1}^{n} w_i = 1 \tag{29}$$

Therefore, the predicted traffic flow at $s_0$ is given by:

$$\hat{q}(s_0) = \sum_{i=1}^{n} w_i q(s_i) \tag{30}$$

The expected value of the predicted traffic flow is equal to the expected value of the true traffic flow:

$$\mathbb{E}\left[\hat{q}(s_0)\right] = \bar{q} = \mathbb{E}\left[q(s_0)\right] \tag{31}$$

The kriging weights $w_i$ are determined by solving the kriging system of $n+1$ equations for the weights $w_i$ and Lagrange multiplier $\mu$ is:

$$\begin{pmatrix} \gamma_{11} & \gamma_{12} & \cdots & \gamma_{1n} & 1 \\ \gamma_{21} & \gamma_{22} & \cdots & \gamma_{2n} & 1 \\ \vdots & \vdots & \ddots & \vdots & \vdots \\ \gamma_{n1} & \gamma_{n2} & \cdots & \gamma_{nn} & 1 \\ 1 & 1 & \dots & 1 & 0 \end{pmatrix} \begin{pmatrix} w_1 \\ w_2 \\ \vdots \\ w_n \\ \mu \end{pmatrix} = \begin{pmatrix} \gamma_{10} \\ \gamma_{20} \\ \vdots \\ \gamma_{n0} \\ 1 \end{pmatrix} \tag{32}$$

where $\gamma_{ij}$ is the semivariance between known points $s_i$ and $s_j$, $\gamma_{i0}$ represents the semivariance between $s_i$ and $s_0$.

To improve reproducibility and clarity, Algorithm 1 outlines the complete workflow for spatial imputation using a variogram-based kriging approach. In Step 1, we begin by calculating the empirical variogram. This involves defining sensor's location, as $\{x_i, y_i\}$ coordinate, and observed traffic flow represented as, $q_i$. From the provided network link information, one can estimate the shortest path distance ($h$) between any two sensor coordinates. For each distance bin $h$, we compute the semivariance $\Gamma(h)$. In Step 2, we fit various theoretical variogram models (for example, Exponential, Spherical, and Gaussian) and calibrate the model parameters, nugget ($C_0$), sill ($C$) for a given reference range ($a$). We evaluate the models based on their goodness of fit to the empirical data and select the model ($\gamma(h)$) that best represents the spatial dependency of the data. The parameters of the chosen model are then determined. Step 3 applies kriging for data imputation. For each missing sensor location, we identify the $k$ nearest known sensors within the range ($a$) defined by the variogram. Kriging weights are computed by solving a system of equations that ensures the weighted sum of the semivariances between known sensors matches the semivariance between known sensors and the missing location. The imputed value at the missing location is then estimated as a weighted sum of the observed values at the known locations.

**Algorithm 1** Spatial Dependency Analysis and Variogram-Based Imputation

**Input:** Sensor data $\mathcal{D} = \{(x_i, y_i, q_i)\}$ with known traffic values
**Output:** Imputed traffic values for missing sensor locations

▷ Step 1: Compute Empirical Variogram

Define shortest-path distance bins: $h \in \{h_1, h_2, \ldots, h_n\}$
**for** each distance bin $h$ **do**

$$\Gamma(h) = \frac{1}{2N(h)} \sum_{i,j \in N(h)} [q(s_i) - q(s_j)]^2$$

Store $(h, \Gamma(h))$ in empirical variogram data
**end for**

▷ Step 2: Fit Theoretical Variogram Models

Fit and evaluate models {*Exponential, Spherical, Gaussian*} to the empirical variogram, $(h, \Gamma(h))$
As an example, the *Spherical* model:

$$\gamma(h) = \begin{cases} C_0 + C\left(\frac{3h}{2a} - \frac{h^3}{2a^3}\right), & \text{if } h \leq a \\ C_0 + C, & \text{if } h > a \end{cases}$$

Select best-fit model and parameters: *nugget* $(C_0)$, *sill* $(C)$, *range* $(a)$

▷ Step 3: Apply Kriging for Data Imputation

**for** each missing sensor location $s_0$ **do**
Identify $k$ nearest known sensor locations $s_i$ within the variogram's range
Compute kriging weights $w_i$ by solving:

$$\sum w_i \gamma(s_i, s_j) = \gamma(s_i, s_0), \quad \forall i, j$$

Estimate missing value:

$$\hat{q}(s_0) = \sum w_i q(s_i)$$

**end for**

## 3. Data

This section describes the empirical data used in this study, focusing on LD data collected under varying levels of sensor deployment. Additionally, it outlines the link hierarchy classification, which forms the basis for the hierarchical scaling methodology proposed in this study.

### 3.1. Data description

The study focuses on the road network of downtown Athens, Greece. The network, excluding minor roads, extends over 150 km in total length (see in Fig. 2). The major roads, characterized by more than two lanes and a peak-hour traffic flow exceeding 1000 vehicles per lane per hour, are identified as the most important and busiest routes within the network. The data for this study was collected from loop detectors over a weekday period from November 7th to 11th and November 14th to 18th, 2022, covering a 24-h time span each day. The loop detector data provides location-based traffic information, such as traffic counts and average speeds.

The links in the road network were classified in two ways. The first classification follows a three-hierarchy (3-H), where roads are categorized into three groups: Link-1, corresponding to the most critical roads, and Link-3, the least important major roads. The second classification follows a two-tier hierarchy (2-H), where the network is divided into two types: Link-1, representing the most important roads, and Link-2, which includes the remaining major roads. Notably, Link-1 is identical in both classifications, while Link-2 in the 2-H classification corresponds to Link-2 and Link-3 in the 3-H classification based on their average speeds and flow profiles. It is important to note that we do not define the 3-H classification ourselves; rather, we adopt the methodology commonly used by mapping service providers such as OpenStreetMap (Haklay and Weber, 2008) and HERE Technologies (see more details in Section 3.2). These classification schemes are widely standardized and readily available for most urban networks globally. No additional link classes beyond the 3-H level were identified as exhibiting statistically significant differences in traffic and road characteristics. Fig. 3 shows the link hierarchies and LDs positions, while Fig. 4 compares average speeds and flow profiles for the 2-H and 3-H classifications. We can see in Fig. 4 that the average speed and flow in each group are significantly different, so using a common flow average from the equipped network to estimate the average flow on the non-equipped network leads to strong bias.

The study focuses on a road network comprising 2456 links, spanning various hierarchical levels within a 150 km length (see in Fig. 2(b)). A fully equipped network is defined as one where every link is equipped with at least one LD. However, only 142 of these links currently have LDs installed. Therefore, the LDs percentage in the network is approximately 5.78%. To address this limitation, a partially equipped network was constructed, utilizing the data from these 142 links to estimate MFDs for validation purposes.

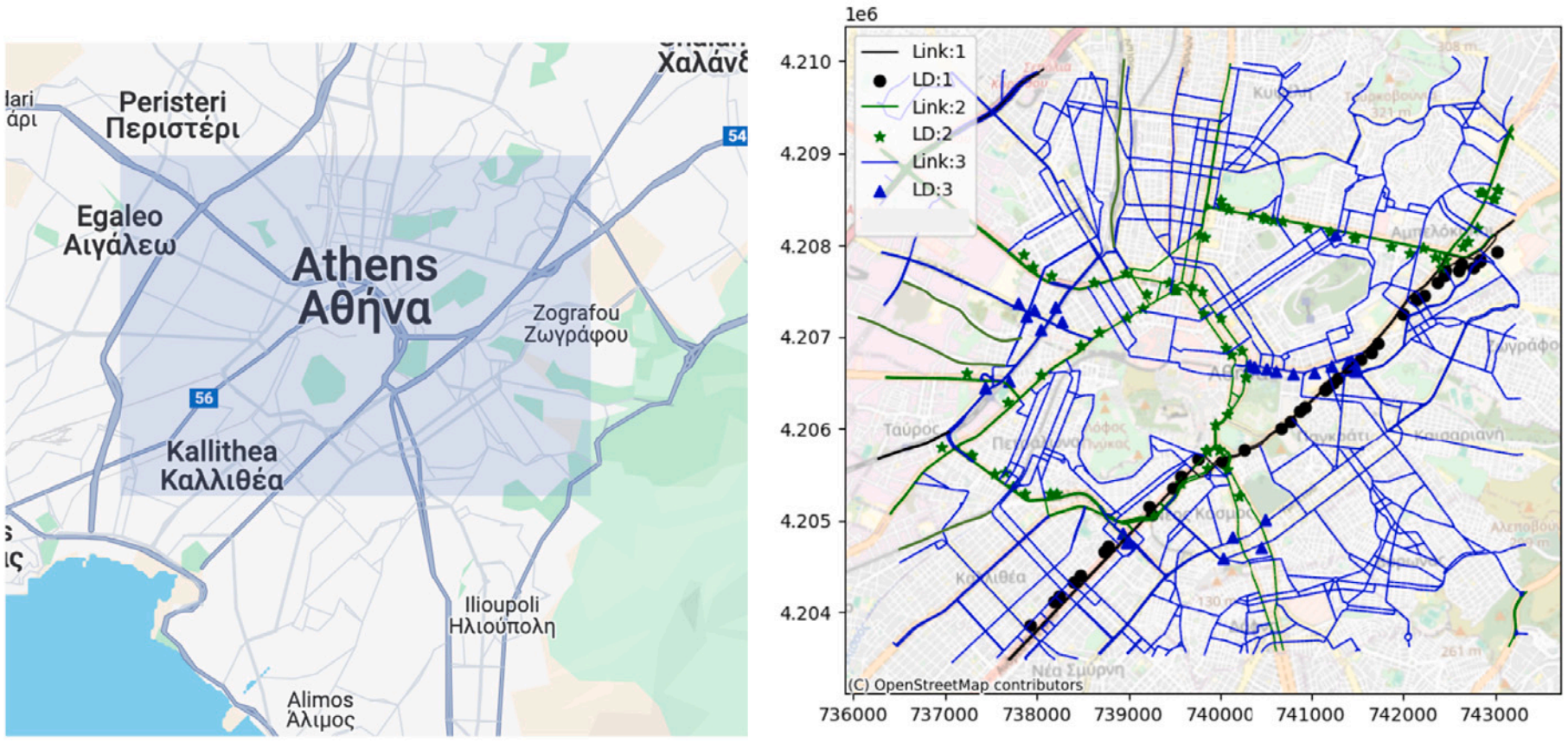

(a) Athens Map (b) Network and LDs

**Fig. 2.** (a) Map view of Athens, Greece, highlighting the study area (source: Google Maps). (b) Full road network with all LDs displayed.

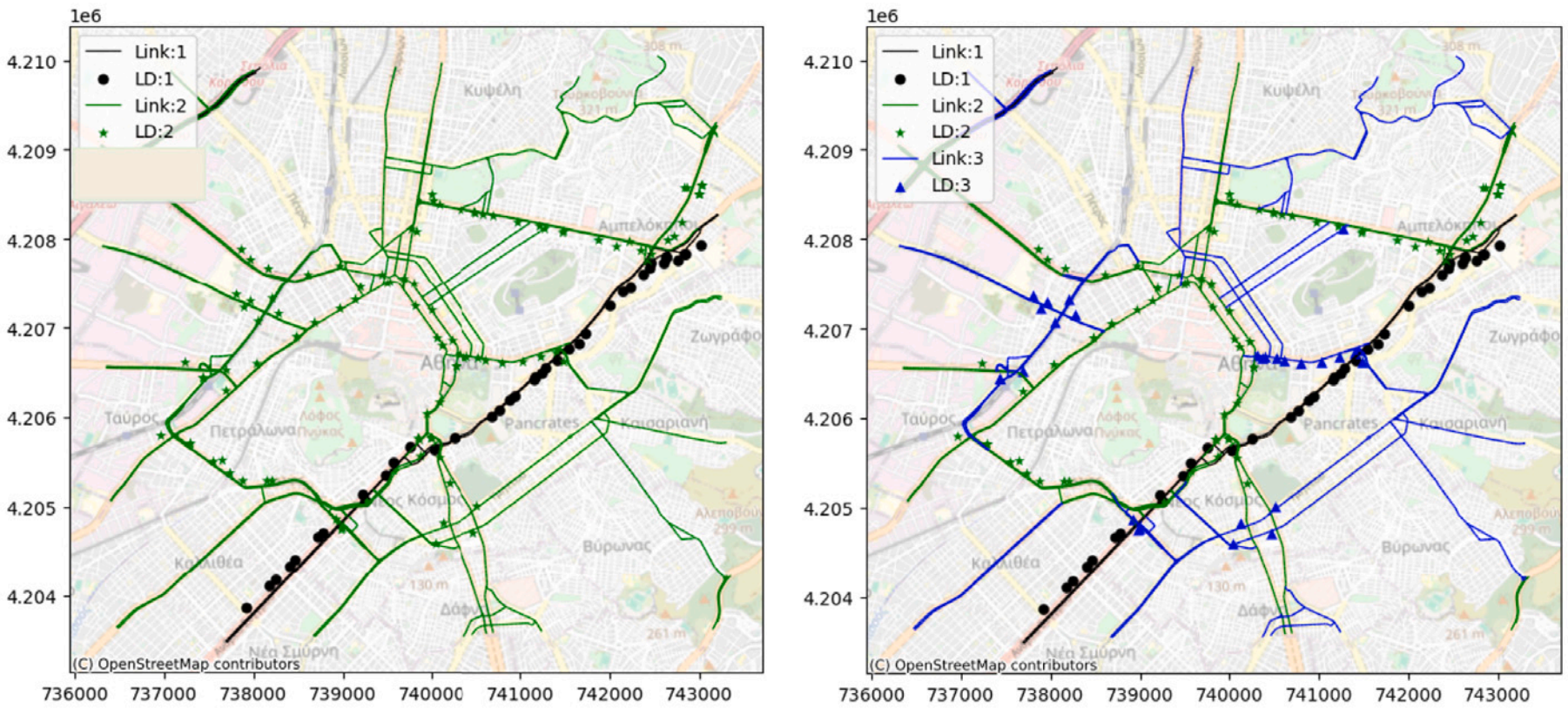

(a) 2-Link Hierarchies (b) 3-Link Hierarchies

**Fig. 3.** Network fully equipped with LDs and features hierarchical links, including (a) two types link, (b) three types link.

**Table 1**
Description of loop detector-equipped link in the fully equipped and partially equipped network.

| Equipped | 3-Hierarchies | | | 2-Hierarchies | | Non-Hierarchy |
|---|---|---|---|---|---|---|
| Network (% LDs) | Link-1 | Link-2 | Link-3 | Link-1 | Link-2 | Link-all |
| 100 | 39 | 75 | 28 | 45 | 95 | 142 |
| 30 | 12 | 22 | 8 | 16 | 26 | 42 |
| 20 | 8 | 15 | 6 | 10 | 19 | 29 |
| 10 | 4 | 8 | 3 | 6 | 9 | 15 |
| 5 | 2 | 4 | 1 | 3 | 4 | 7 |

For comparison, the fully equipped network was conceptualized as including the 142 LDs distributed across their respective links, as illustrated in Fig. 3(a, b). To evaluate our methodology, we also created partially equipped networks by randomly removing LDs from various links equally from each hierarchy in both the 2-H and 3-H hierarchical networks. Table 1 provides details on the number of LDs in each link type for the fully equipped (100%) network as well as for networks with 30%, 20%, 10%, and 5% LDs coverage. Fig. 5 illustrates the distribution of LDs in these partially equipped networks.

In this study, the proposed approach was validated using two distinct network types: the 2-H network, the 3-H network, and a baseline non-hierarchical network for comparison.

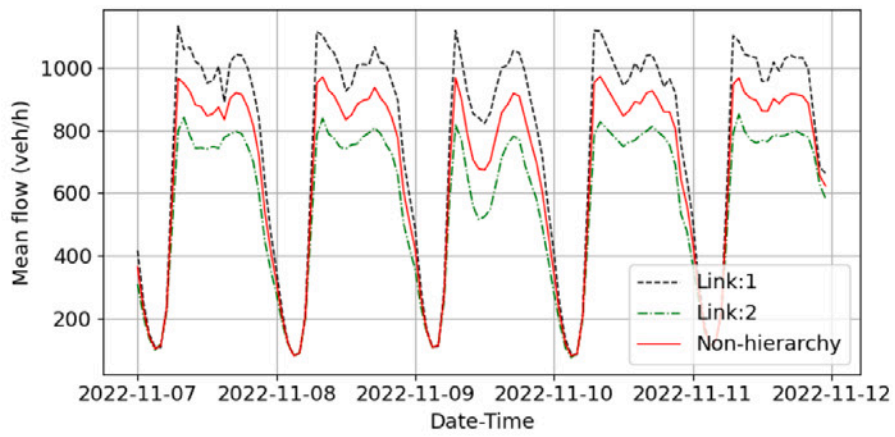

(a) Average flow

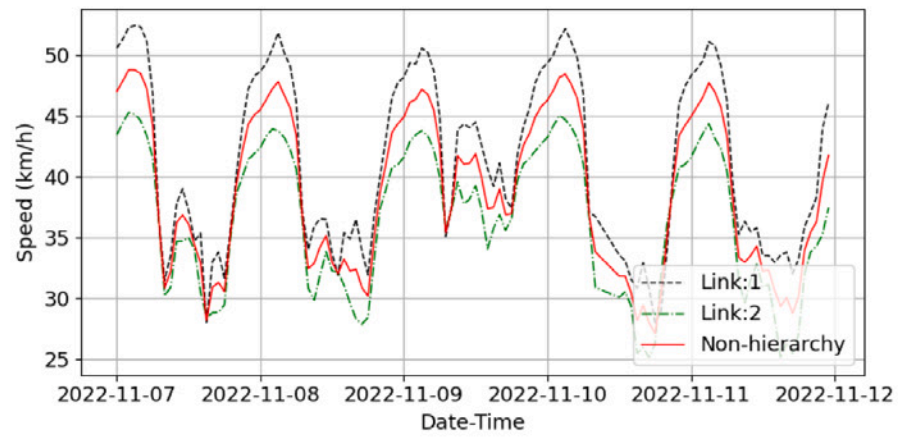

(b) Average speed

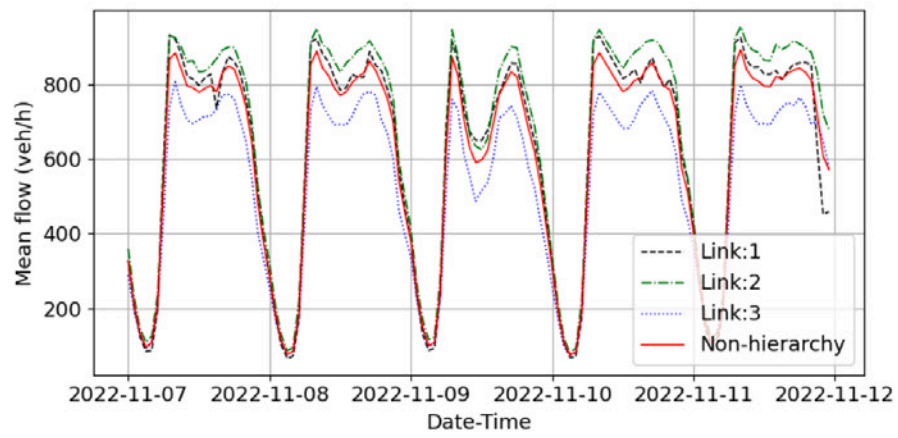

(c) Average flow

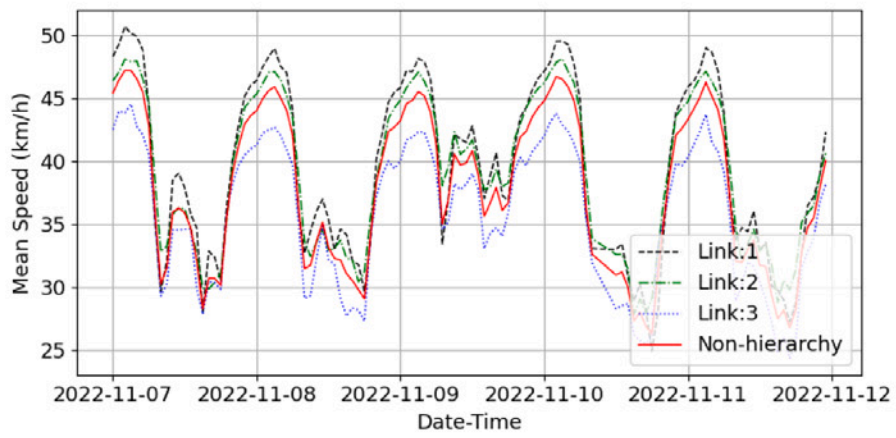

(d) Average speed

**Fig. 4.** Comparison of average network flow ((a) two-hierarchy, (c) three-hierarchy) and speed ((b) two-hierarchy, (d) three-hierarchy) for different network configurations.

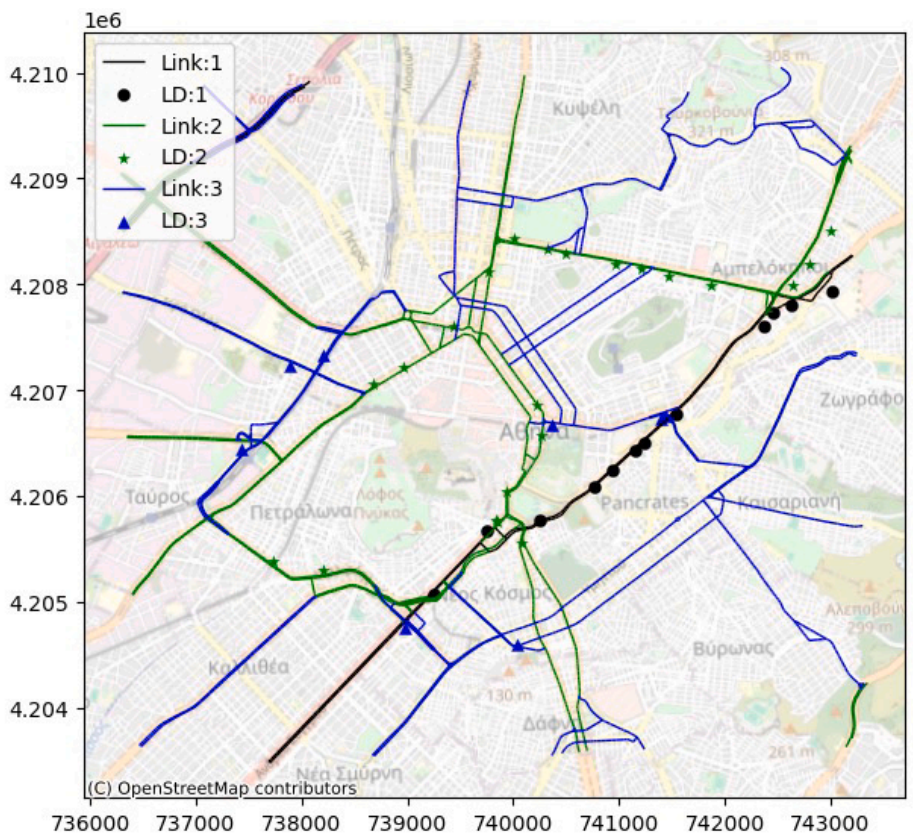

(a) 30% LDs of Figure 3 network

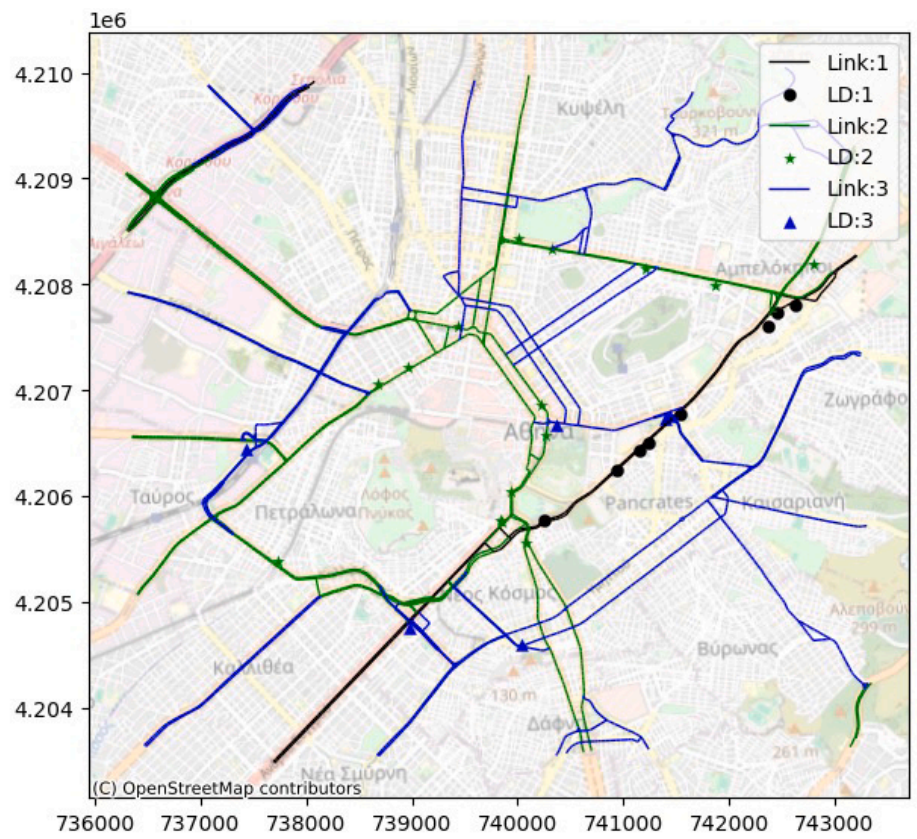

(b) 20% LDs of Figure 3 network

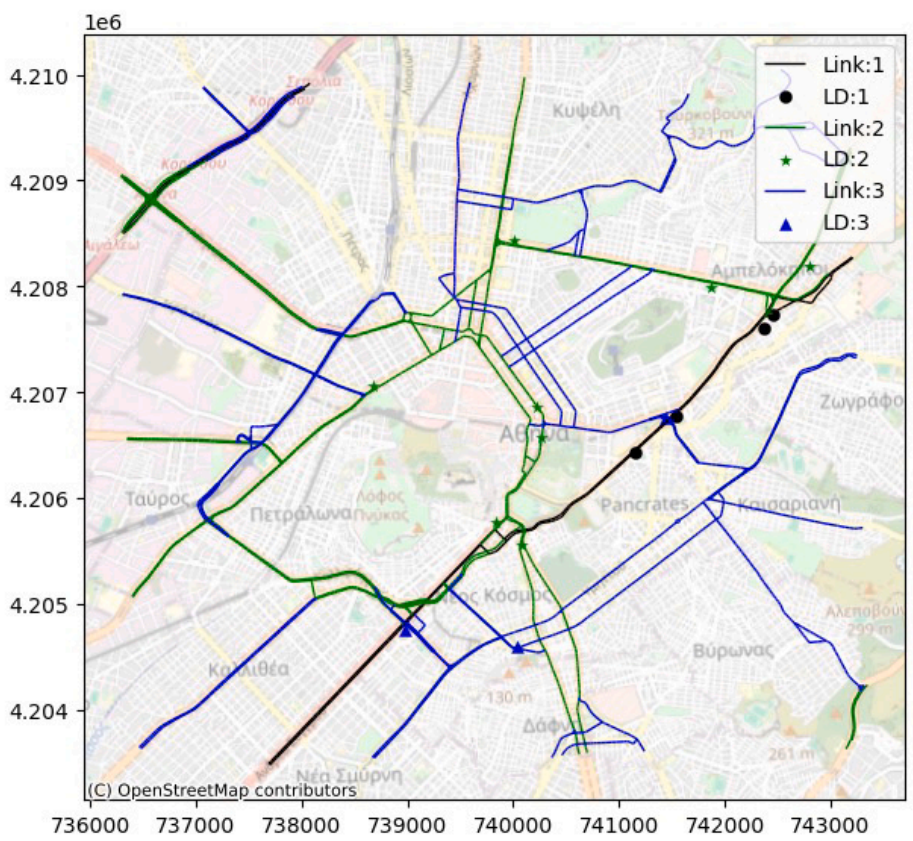

(c) 10% LDs of Figure 3 network

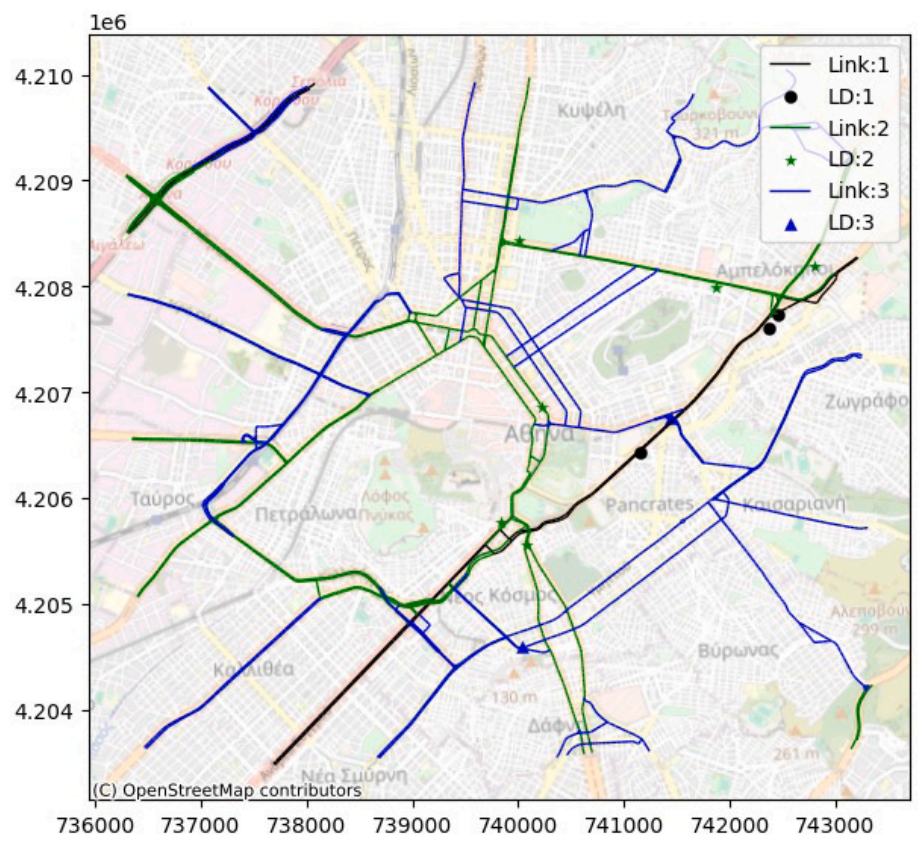

(d) 5% LDs of Figure 3 network

**Fig. 5.** Partially loop detectors equipped networks (a) 30% LDs equipped network, (b) 20% LDs equipped network, (c) 10% LDs equipped network, and (d) 5% LDs equipped network.

### 3.2. Link hierarchy estimation

In this study, we scale traffic variables across non-equipped networks by leveraging a hierarchy-based classification of links. To define these link hierarchies, we adopt a classical approach widely used by mapping service providers such as HERE Technologies and OpenStreetMap. Although more advanced methods, such as data-driven clustering or dynamic partitioning, could potentially refine the hierarchy further, such techniques typically require wide-area sensing coverage. Given the sparse sensor deployment in many urban networks, including the one used in our study, we rely on the established classification schemes, which are both interpretable and readily accessible.

#### 3.2.1. Link Hierarchy with mapping services

Most urban road networks come equipped with standardized metadata accessible from widely used mapping services such as OpenStreetMap (Haklay and Weber, 2008) and HERE Technologies (https://www.here.com/). For instance, HERE provides a *functional road classification* based on the strategic importance, flow levels, and design speed of each road segment. Similarly, OpenStreetMap assigns hierarchical labels such as `motorway_link`, `trunk_link`, `primary_link`, `secondary_link`, and `tertiary_link`, which reflect the functional role of each road within the network. This standardized classification approach has the advantage of being globally available and consistent across cities, making it highly practical for real-world applications where manual classification or rich sensor data are unavailable. Such metadata-driven classification provides a reliable approximation of the relative serviceability of different links. To ensure that the static classifications reflect real traffic behavior, we validate the link hierarchies using observed data such as average travel speed, demand, and their temporal variances over several days. As illustrated in Fig. 4, these indicators exhibit clear and consistent differences across hierarchical levels, affirming the stability of the chosen hierarchy.

#### 3.2.2. Dynamic clustering via network partitioning

While static classifications offer a practical and generalizable starting point, more adaptive clustering approaches can enhance the robustness of link hierarchy estimation, especially in large and heterogeneous networks. Dynamic clustering can be achieved by partitioning the network into homogeneous regions based on real-time traffic conditions such as congestion levels, spatial propagation of delays, or travel time distributions. Several studies have explored dynamic partitioning techniques to improve MFD estimation and traffic control (Saeedmanesh and Geroliminis, 2016, 2017; Saedi et al., 2020; Jiang and Keyvan-Ekbatani, 2023; Jiang et al., 2023). Once such partitions are identified, our method can be applied within each homogeneous region using either static or refined hierarchical classifications. However, it is important to note that such clustering often requires detailed link-level data, which may not be available in sensor-sparse environments.

In this work, we assume that the selected network (or region) is sufficiently homogeneous to define a valid MFD. Within this context, the use of a fixed and readily available link hierarchy is a practical and effective solution for applying our scaling methodology.

## 4. Results

The study proposes two methods: first, hierarchical scaling with application to the 2-H and 3-H network approach. Second, the spatial data imputation with the variogram method to spatially interpolate the network variables in the full network. In the end, we compared the estimated network variables and MFDs with the baseline non-hierarchical network approach.

### 4.1. Flow estimation using spatial imputation methods: variogram

The Variogram model takes input of flow from the equipped network's links and estimates flow spatially over the full network. Fig. 6 demonstrates the flow estimation spatially with respect to different percentages of LD-equipped networks at various times of the day. The color-coded flow plots visually represent traffic flow intensity across the network links, with the gradient indicating varying flow levels, from low (blue) to high (red). Since this study focuses on flow scaling only on the major roads (Link-1, link-2, link-3), we only extracted information along the links in the major roads. In the first column of Fig. 6, where 30% of the network's major links are equipped with LDs, the Variogram model produces a detailed and accurate estimation of flow across all major and minor links. Congestion hotspots, as indicated by red zones, align closely with expected traffic patterns, validating the model's capability to capture traffic dynamics effectively under sufficient LD coverage. As the LDs percentage is reduced to 20% and 10%, shown in the subsequent columns, the model maintains a good estimation quality of the overall network flow, with significant flow details retained for major road links (Link-1, Link-2, and Link-3). Congestion hotspots and the broader distribution of flow are generally well-captured, indicating the robustness of the Variogram-based estimation even under reduced LD coverage.

As we further decreased the LDs percentage to 5%, this method was not able to estimate full network traffic flow because of the sparse information with respect to the variogram radius of estimation. Therefore, this method is best suited for network flow scaling of an equipped network having more than 10% LDs. However, at the 5% LDs level (final column), the model's performance declines. Sparse detector data creates spatial gaps that hinder accurate network flow estimation. This results in less precise representations of flow across the network, with notable discrepancies in congestion hotspots and overall flow intensity. This limitation highlights the importance of maintaining a critical mass of LD coverage for accurate spatial flow estimation. Note that the same limitations would ultimately apply to more advanced ML-based imputation methods (such as GCNs, GANS, etc.) because of data sparsity. Our goal is not to determine the flow on links but to perform imputation in a way that ensures accurate aggregation when calculating the network's average flow.

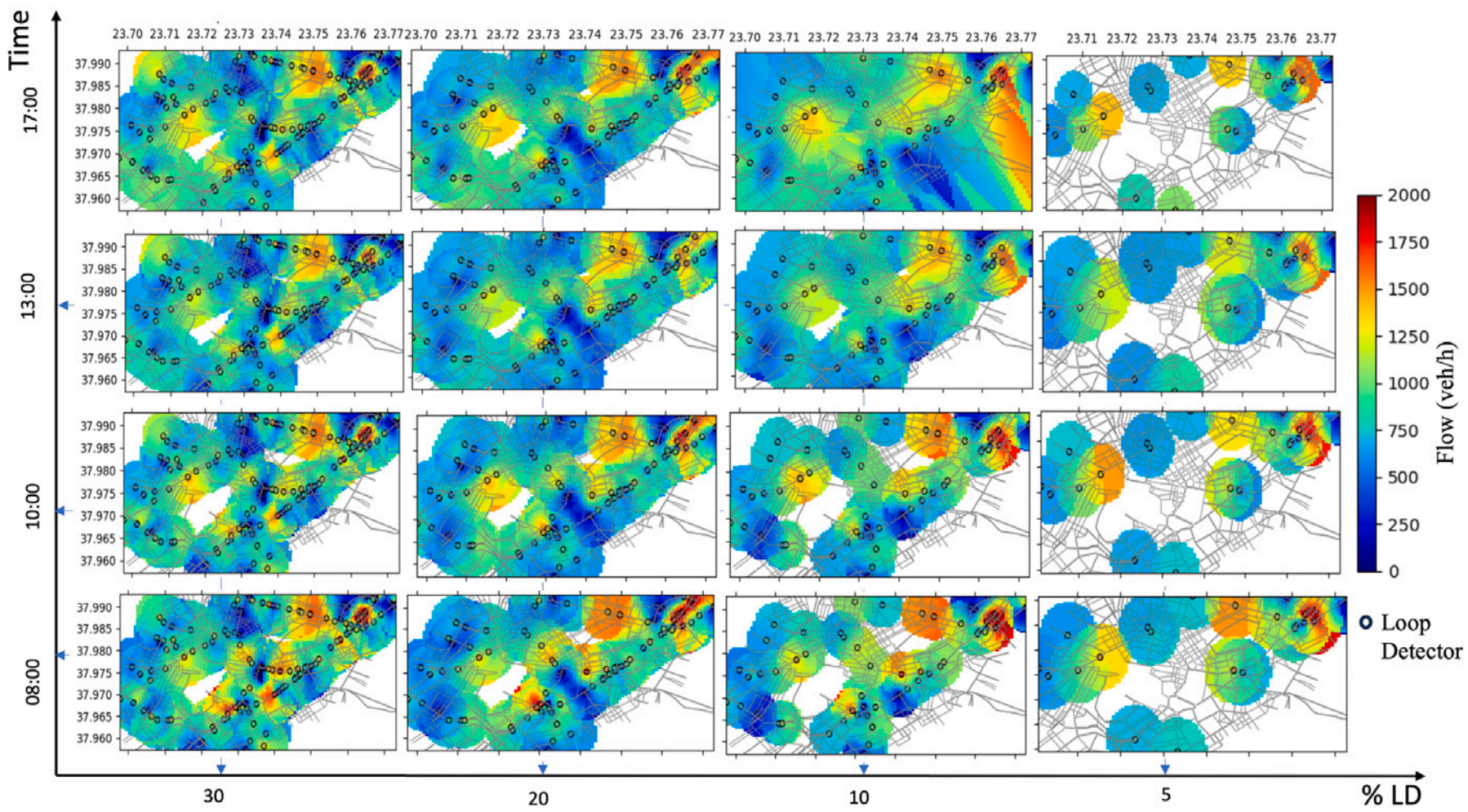

**Fig. 6.** Traffic flow prediction from the equipped network to the entire network using the Variogram method. The *x*-axis represents the equipped network with varying LD percentages, while the *y*-axis depicts traffic flow at different hours of the day at both the LDs position and network level.

Across all LD percentages, flow estimations vary consistently throughout the day (from 08:00 to 17:00). Peak traffic hours show distinct congestion patterns, emphasizing the model's sensitivity to temporal variations in traffic demand and conditions. The consistent ability of the Variogram model to capture these temporal patterns, even as LD coverage changes, is a notable strength. This analysis underscores that the Variogram-based method for network flow scaling performs optimally with LD coverage of at least 10%. Below this threshold, estimation accuracy declines significantly due to sparse spatial data, reducing the reliability of flow predictions.

### 4.2. Comparison of proposed methods

The results presented in Figs. 7 and 8, along with the RMSE values in Table 2, provide a detailed comparison of the proposed hierarchical network scaling method, the variogram-based approach, and the baseline non-hierarchical network scaling. This evaluation focuses on their performance in estimating network flow under different levels of LD coverage, specifically at 5%, 10%, 20%, and 30%. The temporal flow estimations in Fig. 7 show that the baseline (non-hierarchical) quickly loses accuracy when the LD coverage is reduced below 20%. It results in an inaccurate estimation of network flow, causing significant bias in the MFD. Recall the usual network with 5.78% LDs coverage, which is very low and far from 20% LDs coverage. At the lowest LD coverage (5%), the Variogram method is unable to estimate flow across the entire network, reflecting a limitation in this approach when LD coverage falls below 10%. At 10% LDs coverage, the Variogram method becomes viable and achieves relatively low RMSE values compared to non-hierarchical network scaling. Although the variogram method with 10% and 20% LDs coverage demonstrates improved accuracy compared to the baseline, it also results in greater fluctuations in higher flow regimes (see Fig. 7). Meanwhile, At lower LD coverage, the hierarchical network scaling demonstrates superior performance, exhibiting minimal deviation and fluctuation compared to the actual flow. The residual plots in Fig. 7 illustrate the fluctuations of each estimation relative to the actual flow. Notably, the 3-H scaling residuals remain closest to the zero line across all cases, with less fluctuation compared to the other methods. These observations highlight that, among the various methodologies, the 3-H network cluster in the hierarchical approach performs the best overall. The 3-H continues to provide estimations, though with elevated RMSE values (e.g., RMSE of 65–64 on day 1 and day 4 for the 3-Hierarchy method). This highlights the robustness of hierarchical frameworks compared to geospatial imputation techniques like the Variogram, which require a minimum density of LDs for effective network-wide prediction compared to the non-hierarchical scaling approach. Across all methods, higher LD coverage consistently leads to more accurate predictions. RMSE values significantly decrease once LD coverage reaches 20%–30%, a benchmark that is challenging to achieve in practice. The average RMSE for five days of network traffic flow estimation at LD coverage of 10% is 29.2 vehicles per hour (veh/h) using the 3-H approach, compared to 41.6 veh/h with the variogram method. When the LD coverage decreases to 5%, the RMSE for the 3-H method increases to 56 veh/h, which remains a reasonable estimation given the reduced coverage of LDs. In the hierarchical network methods, increasing the number of hierarchy groups in the network consistently enhances estimation accuracy. Therefore, at any given LD coverage percentage, the 3-hierarchy (3-H) network consistently outperforms the baseline and other methods.

The scatter plots in Fig. 8 further illustrate these trends, showing the spread of estimated flow values relative to actual flow values from a fully equipped network. As LD coverage decreases, the spread around the ideal $x = y$ line widens, particularly for the non-hierarchical method, which exhibits greater deviations from actual values. Conversely, the 3-Hierarchy and variogram approaches demonstrate tighter clustering around the ideal line, especially at lower LD coverages, reaffirming their superior accuracy under moderate to high LD densities.

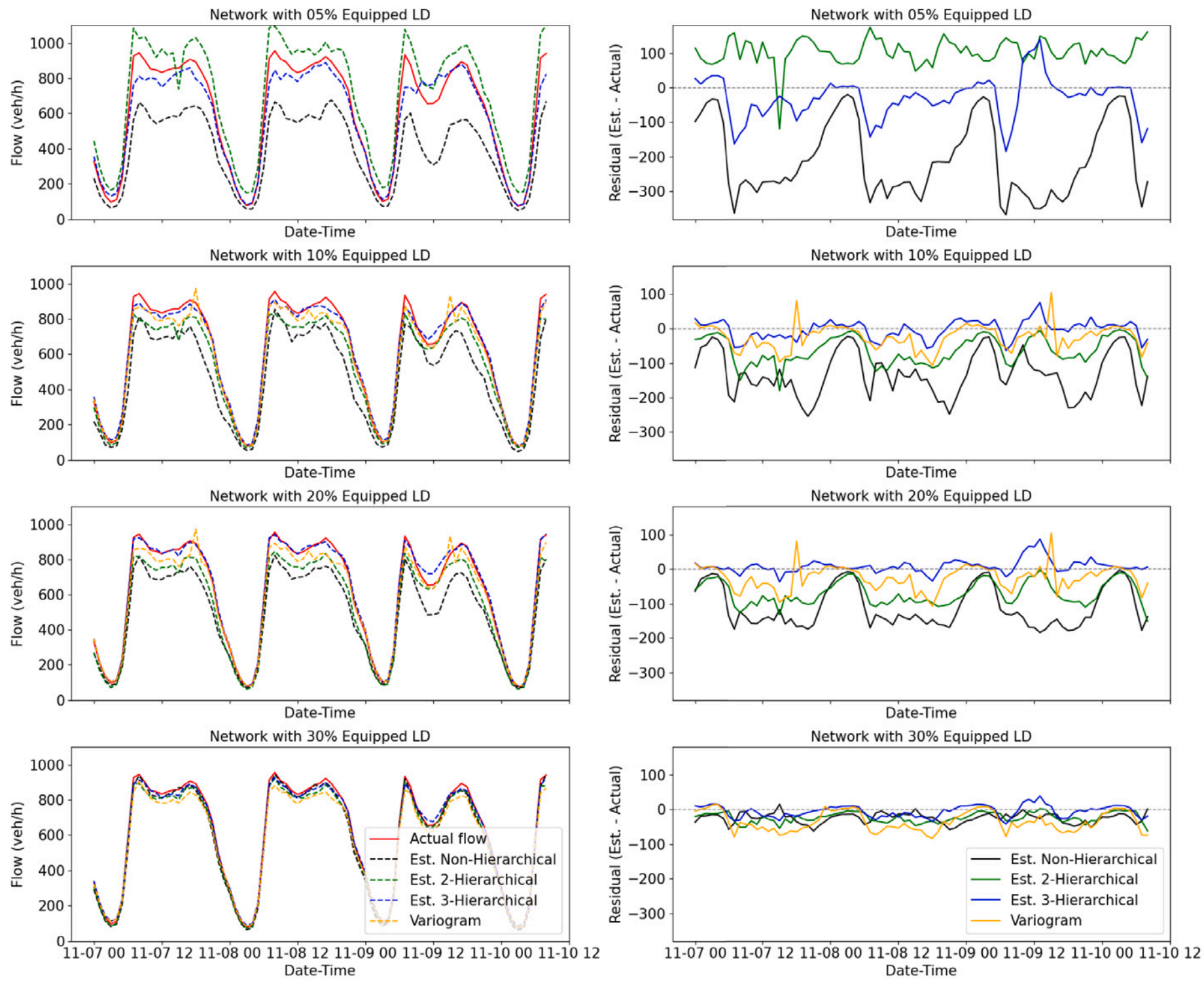

**Fig. 7.** Estimated hourly network flow comparisons using different methods at varying loop detector coverage levels. Left: Actual vs. estimated flows. Right: Residual plots (estimated flow - actual flow).

**Table 2**
Comparison of root mean square error (RMSE) of network flow, estimated using different network scaling methods.

| Day | Network (% LDs) | 3-H | 2-H | Non-hierarchy | Variogram |
|---|---|---|---|---|---|
| 1 | 30 | **12** | 30 | 24 | 12 |
| | 20 | **12** | 85 | 128 | 17 |
| | 10 | **26** | 90 | 158 | 43 |
| | 5 | **65** | 110 | 226 | – |
| 2 | 30 | **13** | 29 | 20 | 15 |
| | 20 | **14** | 83 | 127 | 29 |
| | 10 | **26** | 87 | 148 | 45 |
| | 5 | **53** | 108 | 232 | – |
| 3 | 30 | **25** | 24 | 20 | 42 |
| | 20 | **27** | 70 | 133 | 27 |
| | 10 | **28** | 83 | 142 | 42 |
| | 5 | **51** | 120 | 260 | – |
| 4 | 30 | **12** | 37 | 17 | 25 |
| | 20 | **13** | 94 | 119 | 15 |
| | 10 | **31** | 98 | 144 | 39 |
| | 5 | **64** | 117 | 224 | – |
| 5 | 30 | **13** | 42 | 16 | 31 |
| | 20 | **16** | 94 | 110 | 34 |
| | 10 | **35** | 95 | 145 | 39 |
| | 5 | **48** | 115 | 229 | – |

These results underscore the robustness of hierarchical network scaling, particularly in sparse LD conditions where unstructured methods like the Variogram may falter. The 3-H method, in particular, shows resilience across all coverage levels, while the Variogram method performs strongly at 10% and higher LD coverage, demonstrating its potential for accurate network-wide estimations when sufficient spatial data is available. The baseline approach, applied under the lower usual case of 5% LDs coverage, yields a network-average flow estimation with an RMSE of 234 veh/h. This value is notably four times higher than the error achieved using the proposed three-level hierarchical (3-H) network scaling method. These findings underscore the importance of adopting a hierarchical network approach to improve the accuracy of traffic variable estimation at the network level.

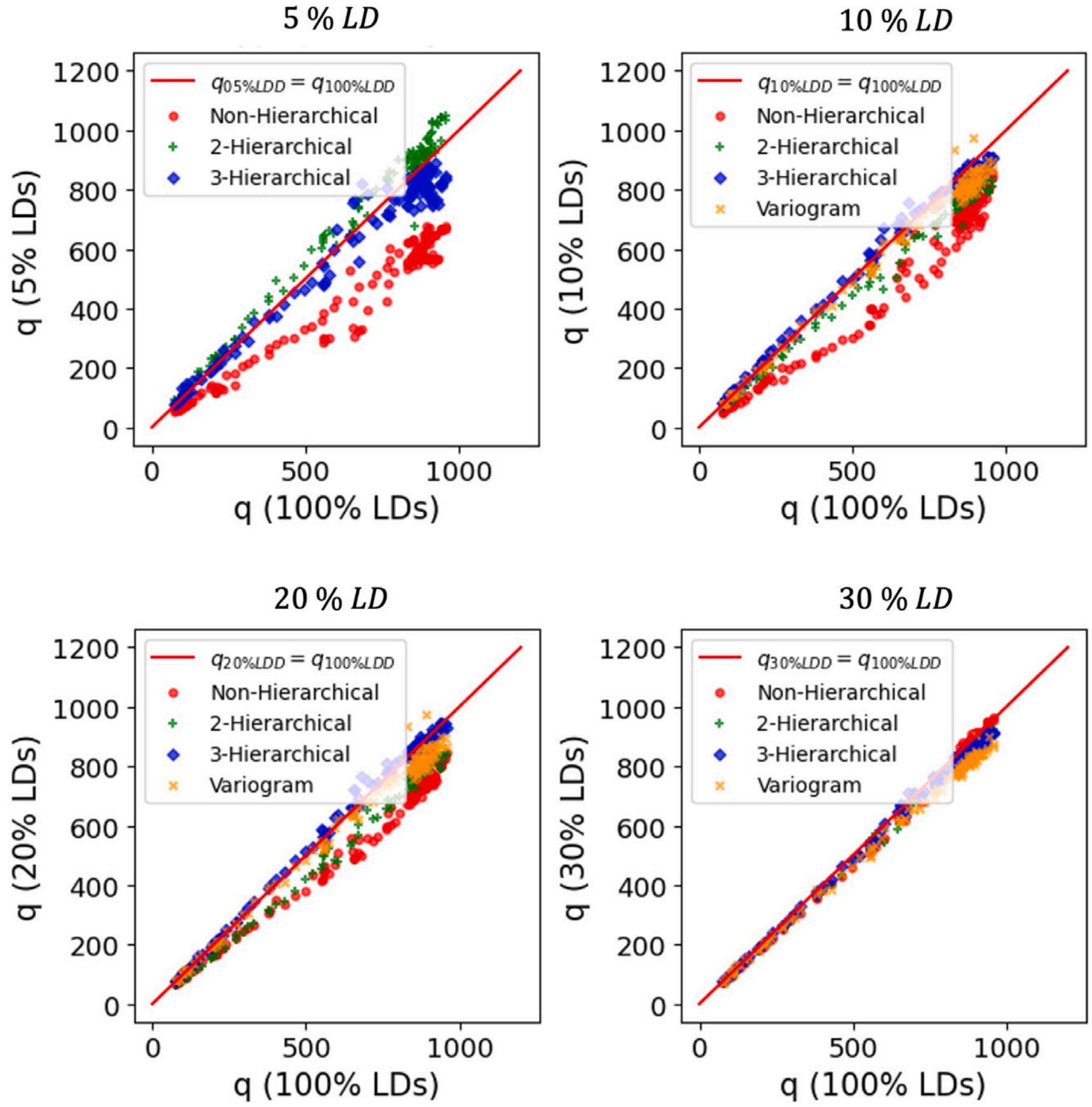

**Fig. 8.** Comparison of actual network flow based on fully equipped network information (100% LDs) with estimated flow derived from a partially equipped network. The $x$-axis represents the actual flow, while the $y$-axis denotes the estimated flow. From left to right, the subfigures illustrate comparisons for 5%, 10%, 20%, and 30% LDs.

### 4.3. Comparing MFDs

This section demonstrates network MFDs estimated from the partially equipped network using the proposed scaling methodologies and compares them with the actual MFDs from the fully equipped network. In Section 2.1, we proposed the network flow ($\hat{q}$) and density ($\hat{k}$) using the hierarchical scaling method, and Section 2.2 described $\hat{q}$ estimation using the variogram. Similarly, we can estimate $\hat{k}$ using variogram methods by estimating detector-level density.

The estimated MFDs derived from the proposed methodologies across varying LD coverage percentages reveal notable differences in estimation accuracy, as detailed in Fig. 9 and Table 3. Under the most data-scarce condition (5% LDs), the three-hierarchy (3-H) method achieves the lowest RMSE and MAPE (48.9 veh/h and 5.2%, respectively), significantly outperforming the non-hierarchical baseline (175.5 veh/h and 17.9%). The two-hierarchy (2-H) and variogram-based methods also yield improved accuracy compared to the non-hierarchical case, though slightly behind 3-H. At 10% and 20% LDs, the hierarchical methods continue to show strong performance, with RMSE and MAPE values generally below 50 veh/h and 5%, respectively. The variogram approach performs comparably, particularly at moderate detector coverage. At 30% LDs, all methods achieve high accuracy, with RMSEs below 40 veh/h and MAPEs near or below 5%. These results emphasize the advantage of incorporating more spatial hierarchy in enhancing the robustness and reliability of MFD estimation, especially under limited data availability.

Using the hierarchical network clustering approach, we estimate the full network MFD, shown in Fig. 10, and compare it with the baseline (uniform scaling) method. The network comprises all 2456 links, as described in the Data Description section. As evident in the figure, the baseline method significantly underestimates the MFD in the critical density range, while both methods perform similarly in the free-flow regime. This observation aligns with the MFDs shown in Fig. 9(a) for 5% LDs coverage, which corresponds closely to the 5.68% coverage in the full network. To support our analysis, we fit the scattered MFD data using a second-degree polynomial and include 95% confidence intervals. Fig. 10(a,b) presents the flow and speed MFDs for Period 1 (07–12 Nov 2022),

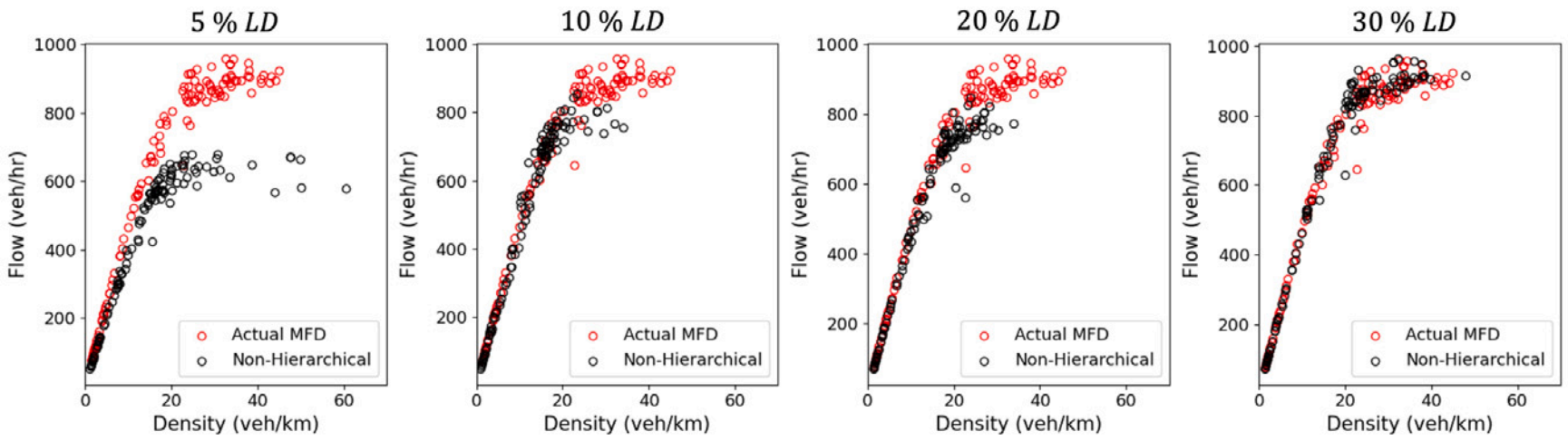

(a) Statistical scaling in non-hierarchical network

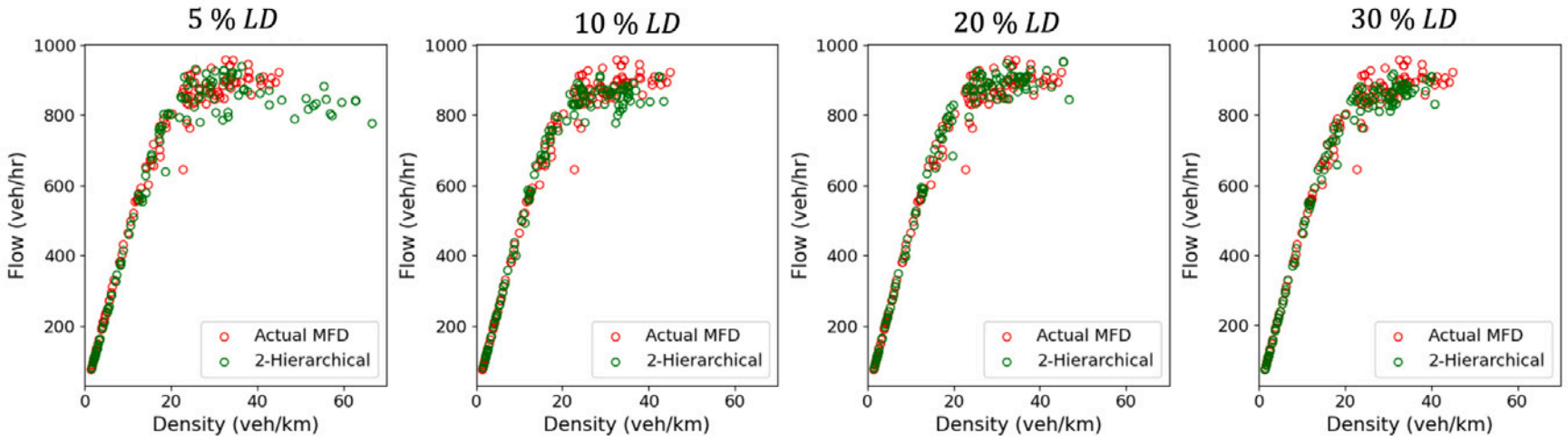

(b) Statistical scaling in two-hierarchical network

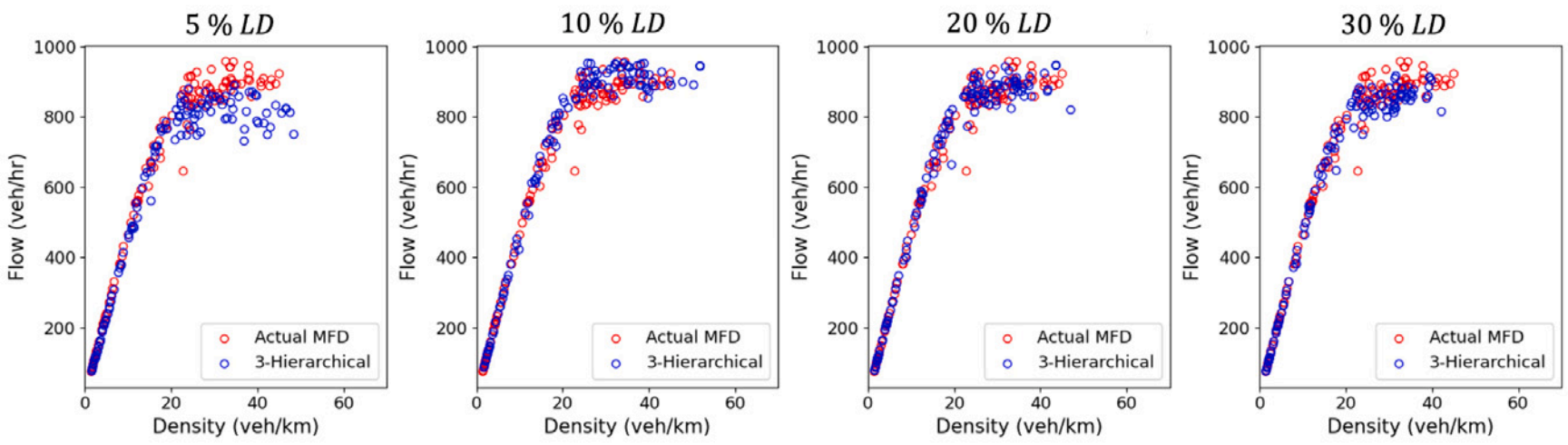

(c) Statistical scaling in three-hierarchical network

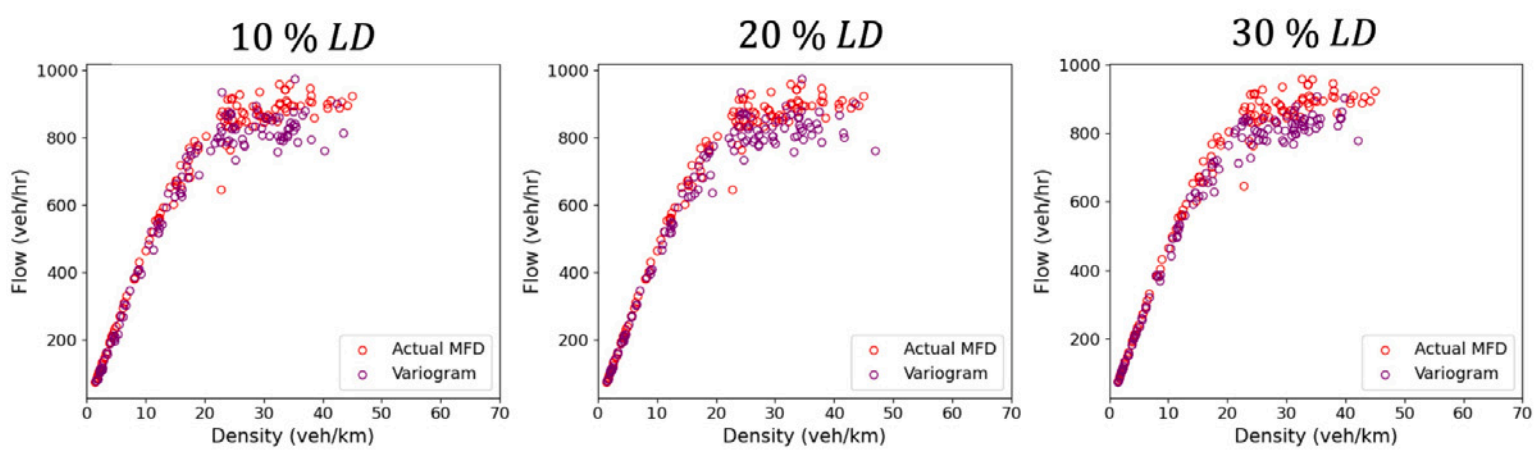

(d) Variogram scaling in non-hierarchical network

**Fig. 9.** Comparison of estimated MFDs from partially equipped networks: (a, b, c) represent hierarchical scaling of 5%, 10%, 20%, and 30% LDs from left to right, and (d) shows variogram scaling of 10%, 20%, and 30% LDs from left to right, compared with the actual MFD derived from a fully equipped (100% LDs) network.

while Fig. 10(c,d) shows results for Period 2 (14–18 Nov 2022), highlighting the impact of the hierarchical scaling method. In the critical density regime, the three-hierarchy model estimates approximately 66 veh/h more flow for Period 1 and around 100 veh/h more for Period 2 compared to the baseline. Additionally, the average network speed improves by about 5 km/h at a density of 30 veh/km, as illustrated in Fig. 10(b,d).

Although the differences in estimated flow and speed at the full-network level are moderate in our case, the impact of hierarchical scaling becomes more pronounced in larger and more structurally diverse networks. In such cases, disregarding network hierarchy can lead to substantial inaccuracies in MFD estimation. Despite the challenge of limited LD coverage in validating the full-network MFD, the proposed hierarchical scaling approach shows clear advantages over the baseline method, particularly in accurately capturing traffic behavior in the critical density regime.

To assess the real-world impact of the proposed hierarchical scaling method, we compare the overall network speed estimation using the speed-MFDs derived from three hierarchical and non-hierarchical scaling methods. Furthermore, we evaluate the accuracy of these estimates over five consecutive days (07–12 Nov. 2022), covering both weekdays and weekends, by comparing them with ground truth probe vehicle speed data. Since probe vehicles traverse the entire network, they provide a reliable approximation of network-wide speed, a method widely validated in previous studies (Gayah and Daganzo, 2011; Geroliminis and Sun, 2011b; Mahmassani et al., 2013; Tsubota et al., 2014; Ambühl et al., 2017; Dakic and Menendez, 2018; Mariotte et al., 2020a). The comparison presented in Fig. 11 demonstrates that the three hierarchical scaling methods yield more accurate network speed estimates, achieving a lower RMSE of 1.89 km/h (MAPE 4.93 km/h) compared to the baseline non-hierarchical method, which results in an RMSE of 2.89 km/h (MAPE 8.22 km/h). This finding highlights the practical advantage of hierarchical scaling in improving network-wide traffic state estimation, making it more suitable for real-world traffic management applications.

**Table 3**
Comparison of estimated and actual MFDs for Athens network.

| Method | Network | % LDs | RMSE (veh/h) | MAE (veh/h) | MAPE (%) | $R^2$ |
|---|---|---|---|---|---|---|
| Hierarchical | Non hierarchy | 30 | 39.9 | 27.0 | 4.96 | 0.98 |
| | | 20 | 133.5 | 90.4 | 11.7 | 0.79 |
| | | 10 | 137.6 | 84.3 | 11.9 | 0.78 |
| | | 5 | 175.5 | 129.8 | 17.9 | 0.64 |
| | Two hierarchy | 30 | 37.1 | 26.9 | 4.2 | 0.98 |
| | | 20 | 37.3 | 27.5 | 4.9 | 0.98 |
| | | 10 | 45.7 | 35.1 | 6.8 | 0.97 |
| | | 5 | 52.0 | 43.8 | 10.4 | 0.96 |
| | Three Hierarchy | 30 | **35.9** | **25.5** | **4.2** | **0.98** |
| | | 20 | **36.5** | **27.0** | **4.5** | **0.98** |
| | | 10 | **45.3** | **29.4** | **4.6** | **0.97** |
| | | 5 | **48.9** | **35.8** | **5.2** | **0.96** |
| Variogram | Non Hierarchy | 30 | 36.4 | 25.2 | 4.2 | 0.97 |
| | | 20 | 38.4 | 27.1 | 4.6 | 0.96 |
| | | 10 | 51.2 | 31.5 | 4.7 | 0.96 |

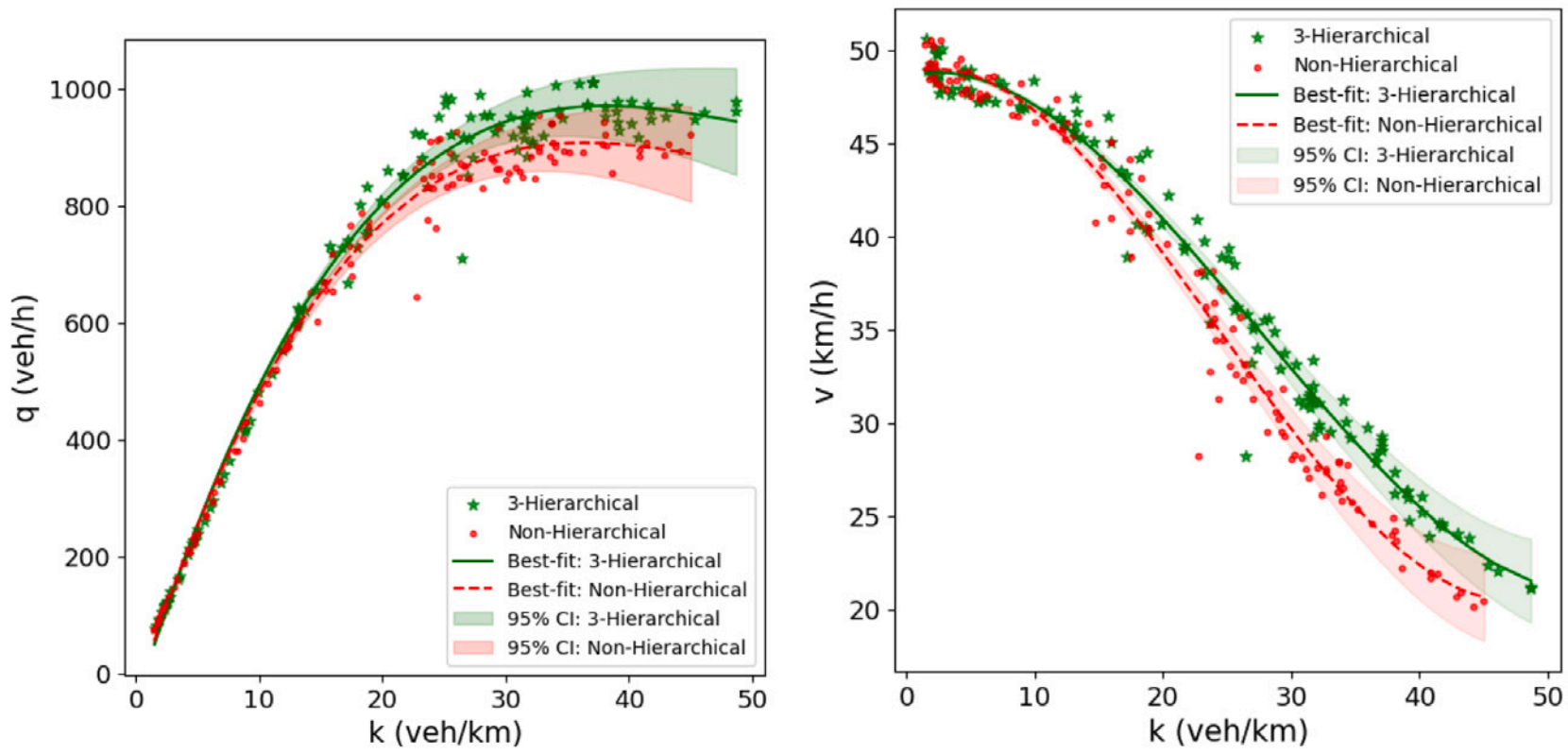

(a) Flow-MFD: 07 - 12 Nov. 2022 (b) Speed-MFD: 07 - 12 Nov. 2022

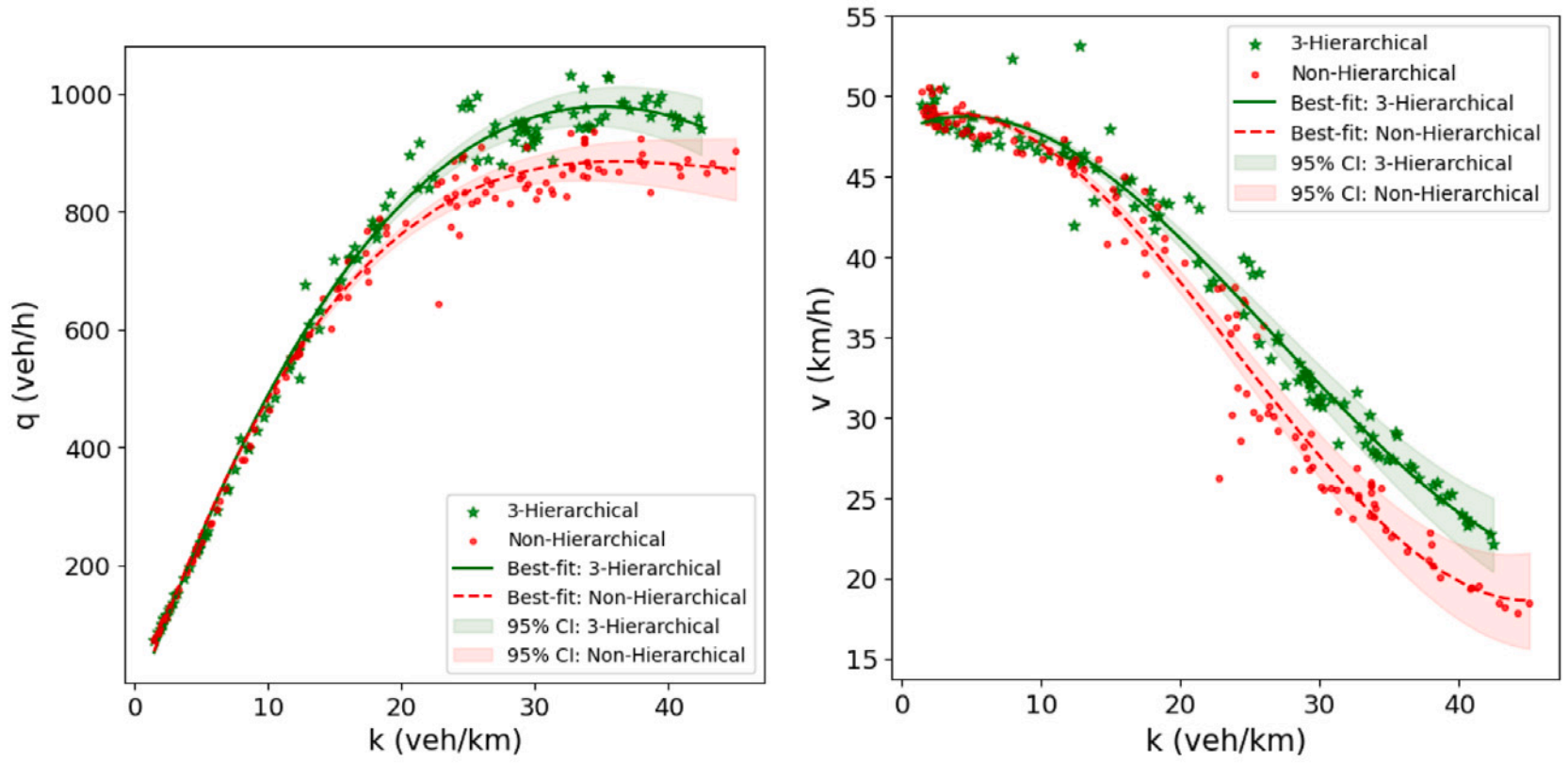

(c) Flow-MFD: 14 - 18 Nov. 2022 (d) Speed-MFD: 14 - 18 Nov. 2022

**Fig. 10.** Comparing the estimated MFD using a hierarchical network cluster with three hierarchies to the baseline non-hierarchical method for the entire Athens network consisting of 2456 links, (a) for the duration of 07–12 Nov 2022, and (b) 14–18 Nov 2022.

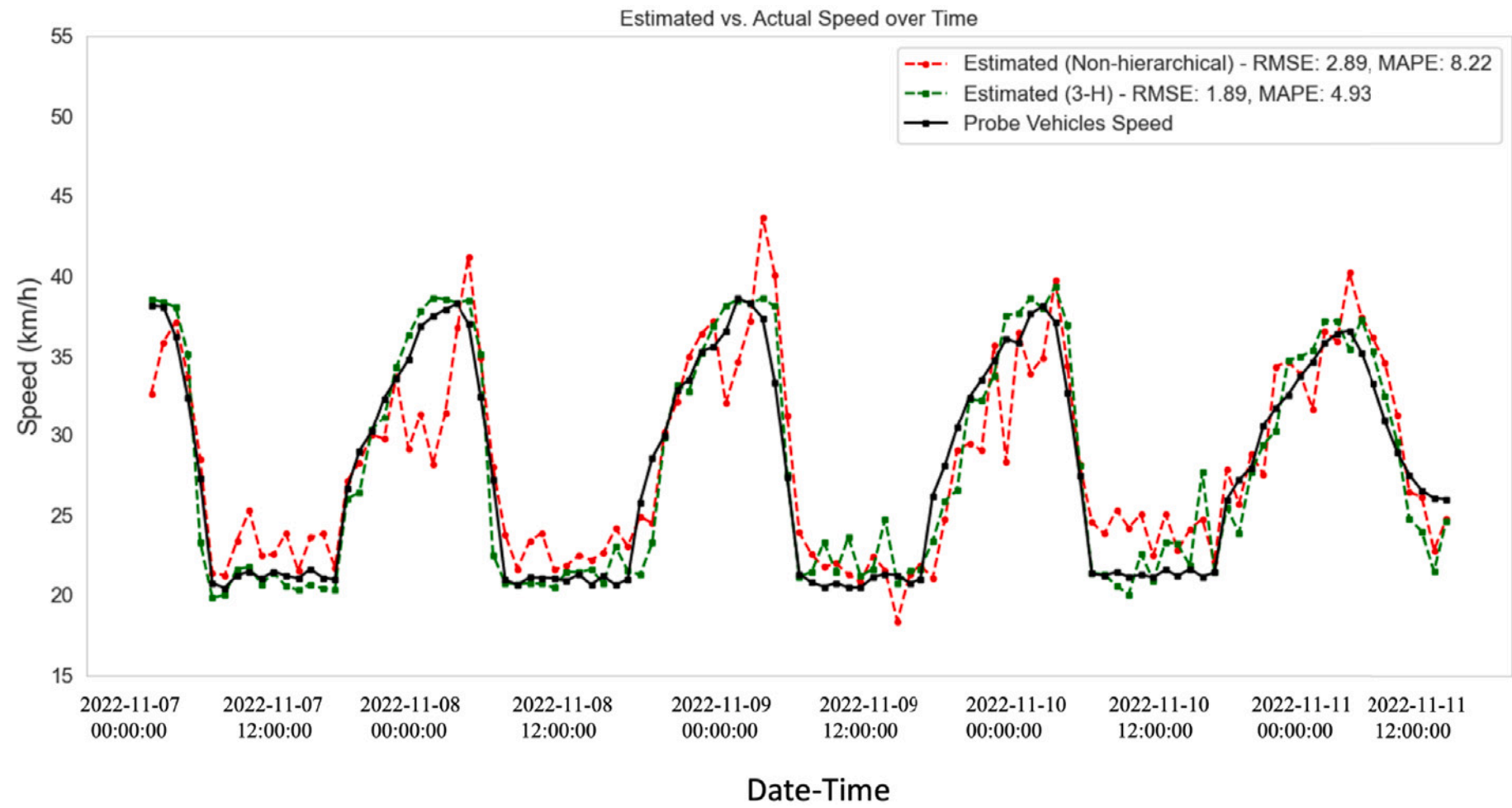

**Fig. 11.** Comparing overall network speed estimation using the speed-MFDs derived from three hierarchical and non-hierarchical scaling methods with the probe vehicles speed for the entire network consisting of 2456 links.

To evaluate whether the differences between the flow-MFDs obtained using hierarchical scaling and those obtained using non-hierarchical (baseline) scaling are statistically significant, we performed a two-tailed paired t-test. This test was designed to detect any significant difference, regardless of whether the estimated flows were higher or lower, between the two methods. The null hypothesis assumed that there is no significant difference in the flow estimates produced by the hierarchical and non-hierarchical approaches, while the alternative hypothesis suggested a statistically meaningful difference between them. The computed t-statistic was 7.0890, which exceeds the critical *t*-value of ±1.9794 at the 95% confidence level. In addition, the *p-value* was less than 0.0001, which is well below the 0.05 threshold for statistical significance. With 123 degrees of freedom, these results lead us to reject the null hypothesis, confirming that hierarchical scaling produces flow estimates that are statistically distinct from those obtained using the non-hierarchical method.

### 4.4. Generalizability and applicability with other network

To evaluate the applicability of the proposed methodology beyond the Athens case study, we employed an additional empirical dataset from downtown Lyon, France. The study area, illustrated in Fig. 12(a), spans approximately 13 km by 20 km and includes data from 238 LDs. These LDs were sourced from an open-access platform (https://avatar.cerema.fr) and cover the period from August 1 to August 7, 2018.

Fig. 12(b) displays the spatial distribution of the detectors and the network layout. The major roads are classified into three hierarchical levels based on functional and geometric characteristics, as described in Table 4. The Link-1 class, representing the highest-level roads in the hierarchy, spans 49 km and includes 658 links, of which 19 are equipped with LDs. The Link-2 category, which forms the intermediate hierarchy, covers 146 km over 3151 links, with 139 links having LDs. Lastly, the Link-3 class, consisting of lower-priority roads, covers 122 km across 2950 links, with LDs present on 80 of them. Altogether, 238 detectors are distributed over 6759 links, resulting in a loop detector penetration rate of approximately 3.5% for the entire network.

To assess the generalizability of the proposed hierarchical scaling method, we applied it to the Lyon network and compared the results, as illustrated in Fig. 13. Despite differing network sizes, spatial configurations, and LD coverage rates (5.68% in Athens vs. 3.5% in Lyon), the hierarchical method consistently demonstrated superior performance over the non-hierarchical baseline. In the flow–density plots, the hierarchical scaling approach estimates significantly higher critical flows: for the Lyon network, an improvement of approximately 100 veh/h at densities around 45 veh/km is observed compared to the baseline method, closely aligning with the 66–100 veh/h improvement range observed in Athens across two study periods. In terms of speed estimation, the hierarchical method yields an average improvement of 4–6 km/h at medium traffic densities (25–35 veh/km) in both networks. These gains are not only evident in the best-fit curves using a second-degree polynomial but are also supported by 95% confidence intervals, suggesting more stable and reliable estimates. The consistent improvements across two structurally different networks confirm the robustness, scalability, and broad applicability of the hierarchical clustering approach for reliable MFD estimation under sparse sensor coverage.

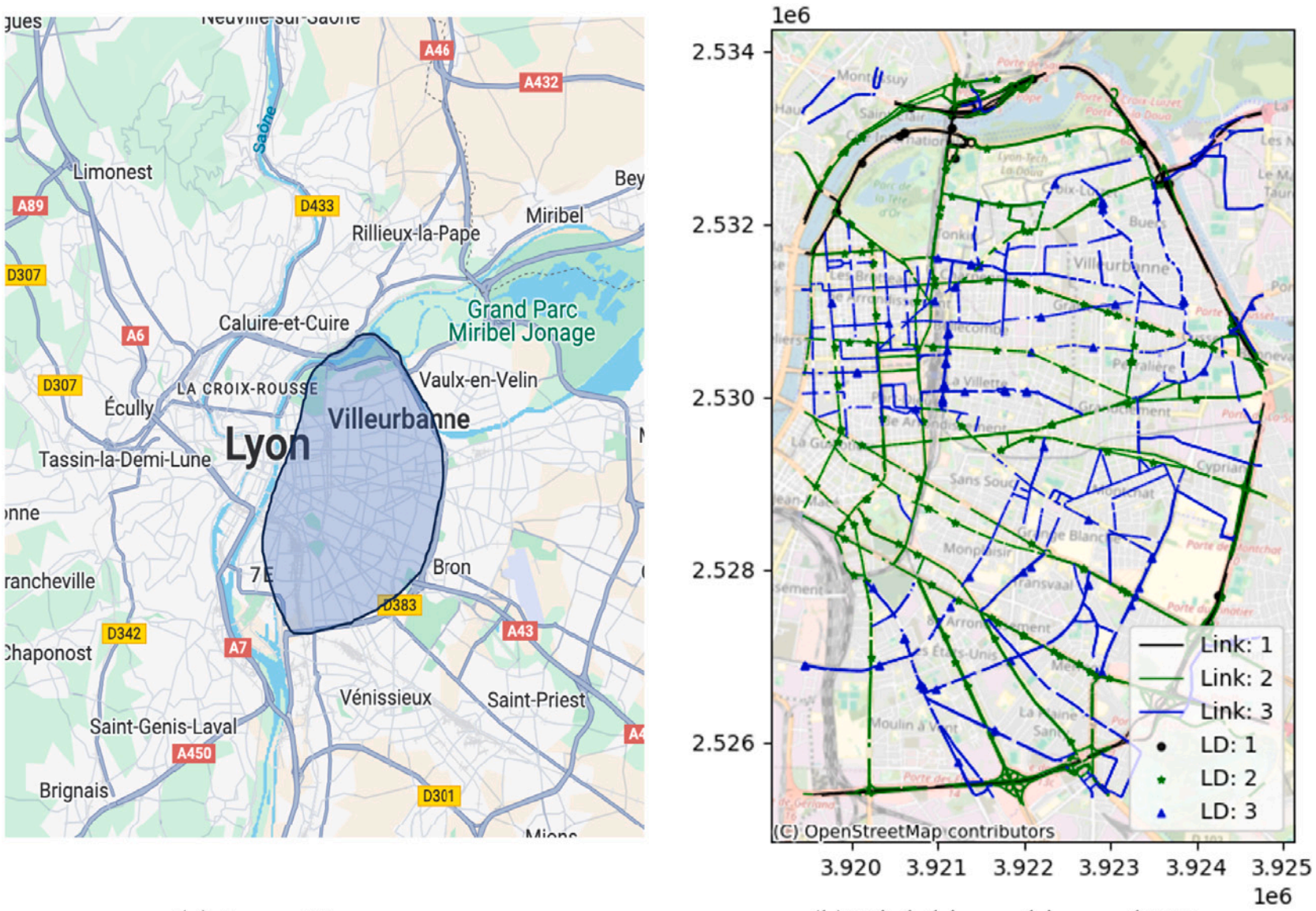

(a) Lyon Map (b) Link hierarchies and LDs

**Fig. 12.** Road network, link hierarchy, and LDs descriptions of Lyon road Network (a) Map view of Villeurbanne, Lyon (source: Google Maps) (b) LDs with corresponding link hierarchies.

**Table 4**
Link hierarchies and LDs information of Lyon network.

| Link hierarchy | Link length (km) | Link number | Detector number | % LDs |
|---|---|---|---|---|
| 1 | 49 | 658 | 19 | 2.9 |
| 2 | 146 | 3151 | 139 | 4.4 |
| 3 | 122 | 2950 | 80 | 2.7 |
| Non-hierarchy | 317 | 6759 | 238 | 3.5 |

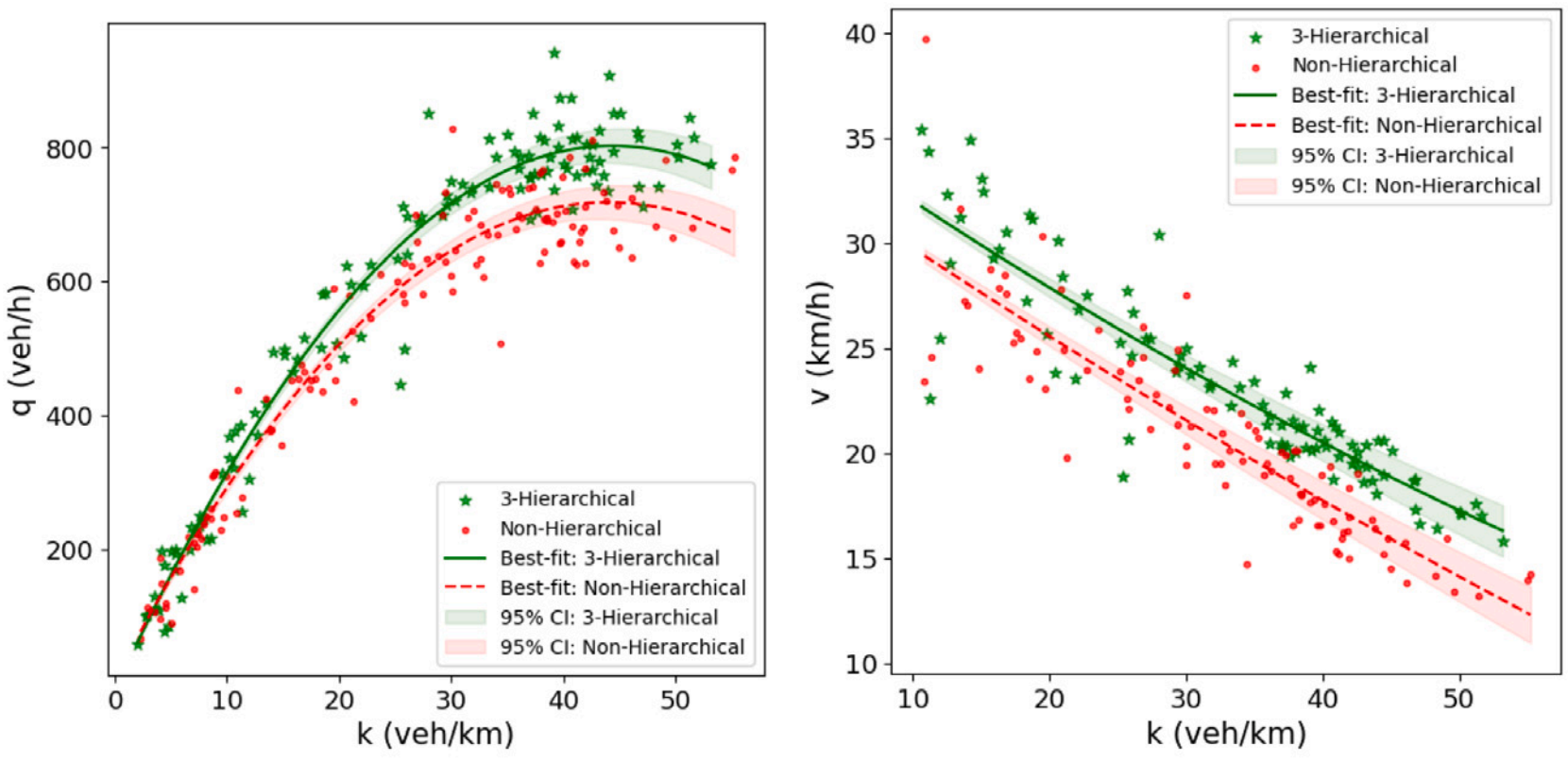

(a) Flow-MFD (b) Speed-MFD

**Fig. 13.** Comparing the estimated MFD using a hierarchical network cluster with three hierarchies to the baseline non-hierarchical method for the entire Lyon network.

## 5. Conclusion

This study introduces a novel methodology for estimating network-wide traffic variables and MFDs in urban environments with limited sensor coverage. By integrating hierarchical network scaling with geospatial imputation via variogram-based kriging, the proposed framework enables accurate estimation of MFDs, even under sparse detector penetration. Our results underscore the effectiveness of hierarchical scaling in improving MFD estimation, particularly in low-data scenarios. Among the tested approaches, the three-level hierarchical method (3-H) consistently outperformed both the two-level (2-H) and variogram-based methods, yielding the lowest RMSE, MAPE, MAE, and highest $R^2$ values across all LD coverage levels. The benefits of the 3-H method were most pronounced under 5% and 10% sensor coverage, where its structured scaling better captured spatial variations in flow and density.

Moreover, our analysis revealed that increasing the number of link hierarchies enhances the accuracy of traffic state estimation and improves MFD representation. For both the Athens and Lyon case studies, the hierarchical approach yielded significantly higher saturation flows and better critical density estimates compared to the baseline method, as shown in Figs. 10 and 13. Although baseline scaling improved with more data, it consistently underperformed in under-monitored regions due to its inability to exploit spatial or structural network properties.

The variogram-based method, on the other hand, demonstrated competitive performance at moderate LD coverage (≥10%), particularly in estimating average speeds and capturing overall flow trends. However, its performance declined at 5% coverage, where insufficient spatial observations limited the reliability of the imputation. This highlights a key trade-off: variogram methods are effective when sufficient spatial data are available, but hierarchical scaling is more robust under sparse conditions. The shortest-path-based variogram model developed in this study offers a computationally efficient and scalable solution for spatial imputation in urban networks. While it simplifies spatial dependence as a function of shortest path distance, this abstraction allows interpolation across the entire network without requiring prior partitioning or hierarchical segmentation. Despite some limitations under extreme data sparsity, the method reliably captures structured spatial dependencies. In contrast, the hierarchical scaling method assumes that links within the same classification share similar mean and covariance properties, allowing for effective extrapolation of traffic conditions. Its simplicity, scalability, and low data requirements make it particularly well-suited for real-world applications in cities with sparse sensor deployments.

While machine learning-based models such as GCNs and GANs have been widely used for traffic state estimation, their application is often limited in scenarios with sparse sensor coverage. These methods require large amounts of labeled data for training and involve considerable computational overhead, making them less practical for real-time or data-scarce environments. In contrast, our proposed hierarchical scaling and variogram-based imputation methods are computationally efficient, interpretable, and specifically designed for urban networks with low sensor density (e.g., ≤5%). They do not require prior training and instead rely on readily available network attributes and spatial relationships.

Overall, our findings emphasize the value of incorporating network structure and spatial dependencies in traffic modeling. The superior performance of the 3-H method across two distinct networks highlights its potential for broad applicability in urban traffic management and planning, especially in settings where full sensor deployment is infeasible. This study thus contributes a practical and transferable solution for bridging the gap between partial sensor data and full-network traffic state estimation.

### *5.1. Scope of applicability*

The proposed methodology is applicable across a wide range of urban network configurations, particularly under the following conditions:

- The hierarchical scaling method is most effective in homogeneously partitioned networks, where clear link hierarchies can be established based on geometric and functional attributes.
- Even in the absence of predefined hierarchies, link classifications can be derived using metadata from sources such as OpenStreetMap or by grouping links by speed limits, lane counts, or demand profiles, etc.
- The variogram-based imputation model does not require any hierarchical classification, making it especially useful in networks where such information is unavailable.
- A minimum number of observed sensor values is required for effective variogram-based interpolation, making its reliability dependent on adequate spatial sensor distribution.
- Both the hierarchical and variogram-based methods are adaptable to a variety of urban environments and can be applied to different network geometries and congestion regimes.
- The framework is particularly suited to stationary sensor-based networks with sparse detector coverage, providing a viable alternative to data-intensive models.

### *5.2. Limitations and future work*

We worked with real test cases using empirical data, which limited our ability to explore scenarios with larger samples of equipped links. Consequently, we were not able to test situations where a more extensive sample of equipped links might be available. We acknowledge that simulation-based benchmarks could have complemented our empirical validation by testing under controlled and more diverse conditions, particularly to assess sensitivity to different network configurations and data availability. However, developing calibrated microsimulation models for these large and complex urban networks would require substantial local field data and intensive calibration efforts, which are beyond the scope of this work. Furthermore, even with microsimulations, accurately reproducing local flow on all links can be challenging due to potential discrepancies between simulated equilibrium and actual driver behavior. Since our methodology relies on realistic link-hierarchy flow patterns to establish reliable scaling across the network, we prioritized the use of empirical data to ensure direct relevance and practical applicability to real-world traffic.

Future research directions include the following:

- The hierarchical method's performance in highly heterogeneous networks with complex link hierarchies remains to be explored. Investigating adaptive network partitioning strategies prior to applying hierarchical scaling could enhance its robustness.
- The variogram model requires careful selection and recalibration of the range parameter depending on the sensor layout of each network.
- Incorporating alternative data sources such as floating car data (FCD), mobile sensing, or crowdsourced traffic observations could help improve accuracy in data-scarce environments.
- Validation using synthetic or simulated networks could offer insight into the method's behavior under higher congestion levels and broader structural diversity.

## CRediT authorship contribution statement

**Nandan Maiti:** Writing – review & editing, Writing – original draft, Visualization, Validation, Software, Methodology, Investigation, Formal analysis, Data curation, Conceptualization. **Manon Seppecher:** Writing – review & editing, Supervision, Project administration, Investigation, Formal analysis, Conceptualization. **Ludovic Leclercq:** Writing – review & editing, Supervision, Resources, Project administration, Methodology, Investigation, Funding acquisition, Conceptualization.

## Acknowledgments

This research has received funding from the European Union's Horizon Europe research and innovation program under Grant Agreement no. 101103808 (ACUMEN). The authors would like to thank Prof. Eleni I. Vlahogianni and her team from NTUA Athens for their support in providing traffic data.

## Data availability

Data: The loop detector data for the Lyon network is publicly available through the CEREMA AVATAR platform (Cerema, 2025). The loop detector data for Athens were provided under a consortium agreement as part of the ACUMEN project, funded by the European Union's Horizon Europe research and innovation programme (Grant Agreement No. 101103808). Due to licensing restrictions, these data are not publicly accessible. Code: The source code used for the automatic calibration of Network Macroscopic Fundamental Diagrams (NMFDs) is available on GitHub at Maiti (2025).